\documentclass[12pt]{article}
\pdfoutput=1

\usepackage{amsmath}
\usepackage{amsfonts}
\usepackage{amssymb}

\usepackage{hyperref}
\newlength{\abstractwidth}
\usepackage{color}
\usepackage{graphicx}

\renewcommand{\thefootnote}{\fnsymbol{footnote}}
\renewcommand{\thanks}[1]{\footnote{#1}}
\newcommand{\starttext}{
\setcounter{footnote}{0}
\renewcommand{\thefootnote}{\arabic{footnote}}}

\newcommand{\bea}{\begin{eqnarray}}
\newcommand{\eea}{\end{eqnarray}}
\newcommand{\ee}{\end{equation}}
\newcommand{\be}{\begin{equation}}

\newcommand{\no}{\nonumber}

\def\Re{{\rm Re}}
\def\Im{{\rm Im}}

\def\no{\nonumber}

\long\def\symbolfootnote[#1]#2{\begingroup%
\def\thefootnote{\fnsymbol{footnote}}\footnote[#1]{#2}\endgroup}

\begin{document}
\starttext
\setcounter{footnote}{0}

\begin{flushright}
IGC-10/10-1\\
\end{flushright}

\bigskip

\begin{center}

{\Large \bf  String Junctions and Holographic Interfaces}

\medskip

\vskip .4in

{\large  Marco Chiodaroli$^{a,d}$, Michael Gutperle$^{a}$, Ling-Yan Hung$^{b}$ and Darya Krym$^{c}$
}

\vskip .2in

{$\ ^{a}$ \sl Department of Physics and Astronomy }\\
{\sl University of California, Los Angeles, CA 90095, USA}\\
{\tt \small  gutperle@physics.ucla.edu}

\vskip .2in

{$\ ^{b}$\sl Perimeter Institute, Waterloo, Ontario N2L 2Y5,  Canada}\\
{\tt  \small jhung@perimeterinstitute.ca}

\vskip .2in

{$\ ^{c}$\sl Instituut voor Theoretische Fysica, Katholieke Universiteit Leuven,\\
Celestijnenlaan 200D B-3001 Leuven, Belgium }\\
{\tt  \small daryakrym@gmail.com}

\vskip .2in

{$\ ^{d}$\sl Institute for Gravitation and the Cosmos, The Pennsylvania State University,\\
University Park, PA 16802, USA }\\
{\tt  \small mchiodar@gravity.psu.edu}

\end{center}

\vskip .2in

\begin{abstract}

\vskip 0.1in

In this paper we study half-BPS type IIB supergravity solutions with multiple $AdS_3\times S^3\times M_4$  asymptotic regions, where $M_4$ is either $T^4$ or $K_3$.
These solutions were first constructed in \cite{Chiodaroli:2009yw} and have geometries given by the warped product
of  $AdS_2 \times S^2 \times M_4 $ over $\Sigma$, where $\Sigma$ is a Riemann surface.
We show that the holographic boundary has the structure of a star graph, i.e. $n$ half-lines joined at a point.
The attractor mechanism and the relation of the solutions to junctions of self-dual strings in six-dimensional supergravity
are discussed.

The solutions of \cite{Chiodaroli:2009yw} are constructed introducing two meromorphic and two harmonic functions defined on $\Sigma$.
We focus our analysis on solutions corresponding to junctions of three different conformal field theories and show that
the conditions for having a solution charged only under Ramond-Ramond three-form fields reduce
to  relations involving the positions of the poles and the residues of the relevant harmonic and meromorphic functions.
The degeneration limit in which some of the poles collide is analyzed in detail.
Finally, we calculate the holographic boundary entropy for a junction of three CFTs
and obtain a simple expression in terms of poles and residues.

\end{abstract}

\baselineskip=16pt
\setcounter{equation}{0}
\setcounter{footnote}{0}

\newpage

{\small

\tableofcontents

}
\newpage

\baselineskip 16pt

\section{Introduction}
\setcounter{equation}{0}

The introduction of conformal boundaries plays an important role in many developments and applications of conformal field theory.
In particular, the work of Cardy \cite{Cardy:1989ir} initiated the project of classifying
all conformal boundary conditions for two-dimensional CFTs. Conformal boundaries are also intensely studied in string theory, where they
provide world-sheet descriptions of D-branes.

Conformal interfaces and defects can be regarded as a generalization of boundary conformal field theories.
In an interface theory, two different conformal theories, $CFT_1$ and $CFT_2$, are separated by a hypersurface of co-dimension one.
The theory preserves a subgroup of the two-dimensional conformal group which leaves the hypersurface invariant.
If additional symmetries, such as current algebras or superconformal symmetries,  are present,
one can also demand that the interface   preserves some subgroup of these symmetries.

The folding trick \cite{Oshikawa:1996dj,Bachas:2001vj} relates a two-dimensional conformal interface between $CFT_1$ and $CFT_2$ to a boundary CFT
in the tensor product $CFT_1\otimes CFT_2$. Using this framework, many interesting questions can be addressed.
The Cardy conditions can be used to classify all possible conformal interfaces  in rational CFTs.
The entanglement entropy between  $CFT_1$ and $CFT_2$ can be used to calculate the $g$-function or boundary entropy \cite{Affleck:1991tk}, which
counts the ground-state degeneracy of the interface.
It is also possible to compute reflection and transmission matrices of the excitations in the bulk CFT,
as well as the Casimir energy  of two interfaces separated by  a finite distance \cite{Bachas:2001vj}.

In condensed matter physics, two-dimensional conformal field theories can be employed to study two-dimensional quantum liquids, while
conformal interfaces  can be used to describe impurities at critical points, quantum wires or to  study the Kondo effect.
Furthermore, a special class of defects which are totally transmissive, the so-called topological interfaces, have been recently investigated
in string theory \cite{Bachas:2007td}.

The AdS/CFT correspondence \cite{Maldacena:1997re,Gubser:1998bc,Witten:1998qj}
offers a powerful tool to study conformal field theories at strong coupling.
A particularly well studied incarnation of the correspondence relates type IIB string theory on
$AdS_3\times S^3\times M_4$ (where $M_4$ is either $K_3$ or $T^4$) with a two-dimensional conformal field theory.
This type IIB background can be obtained by taking the near-horizon limit of  a bound state of $N_1$  D1-branes and $N_5$ D5-branes
 wrapped on $M_4$. The D1/D5 bound state can also be described by the Higgs-branch of the two-dimensional $U(Q_1)\times U(Q_5)$
gauge theory living on the intersection of the branes. In the infrared limit, the theory flows to a ${\cal N}=(4,4)$ two-dimensional
 superconformal theory \cite{Witten:1997yu}. This CFT  can also be understood as a hyperk\"ahler sigma model whose  target  space is $(M_4)^n/S_n$,
 where $S_n$ is the n-dimensional symmetric group \cite{Seiberg:1999xz,Vafa:1995bm,Dijkgraaf:1998gf}.
Another interesting example is given by the holographic realizations of two-dimensional quantum liquids
which have been discussed in \cite{Hung:2009qk}.

An interface or a defect can be realized in the context of the AdS/CFT correspondence by introducing in the $AdS_{d+1}$ bulk spacetime a probe
brane with an $AdS_d$ worldvolume, which is stretched all the way to the boundary \cite{Bachas:2001vj,Karch:2000gx,Aharony:2003qf,Bachas:2002nz}.

An alternative approach is to consider regular supergravity solutions which realize  holographically the reduced conformal symmetry of defect.
In general, these solutions are not globally asymptotic to a single Anti-de Sitter space. Instead,
they have two or more $AdS$ asymptotic regions where the dilaton or some other scalar assume different values.
The first example of such solutions was the Janus solution constructed in \cite{Bak:2003jk} which is a fat dilatonic domain wall
 with $AdS_4$ worldvolume. In recent years, examples of Janus solutions which preserve superconformal symmetries have been found
  in gauged supergravity \cite{Clark:2005te}, type IIB supergravity \cite{D'Hoker:2007xy,D'Hoker:2007xz,D'Hoker:2007fq,D'Hoker:2006uv}
and M-theory \cite{D'Hoker:2008wc,D'Hoker:2009gg,D'Hoker:2008qm,D'Hoker:2009my} (see also \cite{Gomis:2006cu,Lunin:2006xr,Lunin:2007ab,Yamaguchi:2006te,Gomis:2006sb,Kumar:2002wc,Kumar:2003xi,Kumar:2004me,Lunin:2008tf} for related work by other authors on this topic).

Probe branes associated with superconformal defects in $AdS_3\times S^3$ have been discussed in
\cite{Bachas:2002nz,Bachas:2008jd,Raeymaekers:2006np,Yamaguchi:2003ay,Raju:2007uj,Mandal:2007ug}.
 Type IIB supergravity solutions that describe defects and are
locally asymptotic to $AdS_3\times S^3\times M_4$ were  recently  found   in  \cite{Chiodaroli:2009yw}, and
in the present paper we continue the analysis of these solutions.
In particular, we focus our analysis on the so-called \emph{multi-Janus} solutions, which have more than
two asymptotic $AdS_3$ regions and are interpreted as duals of defect theories where more than two CFTs are glued together at a defect.
Simple examples of such junctions of multiple CFTs were considered
in the context of quantum field theories on star graphs \cite{Bellazzini:2006kh}. In these theories, different CFTs propagate freely
on each branch of the graph while the vertex is treated as a defect interacting with the fields and is characterized
by transmission and reflection matrices. CFTs on star graphs have potential applications in the theory of quantum wires.

The organization of the paper is as follows: in Section \ref{sectwo} we review the regular solutions obtained in
\cite{Chiodaroli:2009yw}.
In Section \ref{secthree} we evaluate the Page charges for five-branes and one-branes.
In Section \ref{attracmech} we discuss  the construction of junctions of self-dual strings in six flat dimensions which preserve  a quarter of the supersymmetry.
The behavior  of the scalars in these configurations is analyzed using the attractor equations and it is argued,  by  taking
a near-horizon limit, that they
are related to the  half-BPS solutions of Section \ref{sectwo}. In Section \ref{Three-Jct}, we focus on the case $n=3$, i.e. a junction of three CFTs. We present the conditions on the moduli which set the NS charges to zero in the asymptotic regions and  consider various
degeneration limits of the solution.
In particular,  we analyze a  probe limit where all NS charges vanish and the R-R charges in one
region are much smaller than the charges in the two other regions.
Another interesting class of tractable solutions can be obtained by taking a
doubly-degenerate limit. In this case, all charges can be taken to be of order one.

As an application,  we calculate the holographic boundary entropy for a junction of three CFTs  in Section \ref{holoent}
and compare the result to a toy model CFT calculation  to illustrate some features of
the holographic result. A similar computation for a junction with two conformal field theories was carried out  in \cite{Chiodaroli:2010ur}.
We close with a discussion of our results and directions for future research in Section \ref{discus}.

\section{Review of the Half-BPS interface solutions}\label{sectwo}
\setcounter{equation}{0}

 In this section we review the half-BPS solutions found in \cite{Chiodaroli:2009yw}.  These solutions preserve eight of the sixteen supersymmetries of the Anti-de Sitter vacuum and are locally asymptotic to $AdS_3\times S^3\times M_4$. The metric is given by the warped product of $AdS_2\times S^2 \times M_4$ over a two-dimensional Riemann surface with boundary, $\Sigma$.
$M_4$ is either $K_3$ or $T^4$, and we do not consider fluxes that wrap internal cycles of $M_4$ except its volume.
 The solution contains two scalar fields, i.e. the dilaton, $\phi$, and the axion, $\chi$, the complex three-form, $G$, and the real self-dual five-form, $F_{(5)}$. The complex three-form is a composite of the NS-NS field strength, $H_{3}$, and  the R-R field strength, $F_{3}$.
  \be
 G=e^{-\phi/2} H_{3}+ i e^{\phi/2}\Big(F_{3}-\chi H_{3}\Big)
\ee
 A four-form potential can be defined for $F_{(5)}$. By self-duality, its two components, one along $M_4$ and the other along $AdS_2\times S^2$ are equal up to a contraction with the Levi-Civita tensor.

In \cite{Chiodaroli:2009yw}, it  was shown that the BPS equations and Bianchi identity reduce to the requirement that four combinations of fields and metric factors
are meromorphic or harmonic functions. Hence, the entire solution can be parameterized
 by two meromorphic functions $A(z),B(z)$ and two harmonic functions $H(z,\bar z),K(z,\bar z)$, together with the dual harmonic function
$\tilde K (z, \bar z)$.
All functions depend only on the coordinates $z,\bar z$ of the two-dimensional Riemann  surface  $\Sigma$.

\subsection{Expression for the fields}\label{expfield}
The ten-dimensional metric is given by a fibration of $AdS_2\times S^2\times M_4$ over $\Sigma$. $M_4$ can be $T^4$ or $K_3$,

 \be\label{ansatz}
ds^{2} = f_{1}^{2 } ds^{2}_{AdS_{2}} + f^{2}_{2}ds^{2}_{S^{2}} + f^{2}_{3}ds^{2}_{M_{4}}  + \rho^{2 }dz  d\bar z
\ee
The metric factors associated with $AdS_2$, $S^2$, $M_4$ and $\Sigma$ are given by
\bea
f^2_1 &=&  {e^{ \phi} \over 2 f_3^2} {|H| \over K}  \Big( (A + \bar A) K  -  (B - \bar B)^2 \Big) \label{sol-f1} \\
f^2_2 &=&  {e^{\phi} \over 2 f_3^2} {|H| \over K} \Big(  (A+ \bar A) K -   (B  + \bar B)^2  \Big) \label{sol-f2} \\
f_3^4 &=& 4  { e^{-\phi} K \over A + \bar A}\label{f34formula}\\
\rho^4 &=&  e^{-\phi} K  {|\partial_z H|^4  \over H^2} { A + \bar A \over |B|^4 } \label{sol-rho}
\eea

The dilaton and axion are given by \footnote{In our  previous paper \cite{Chiodaroli:2009yw}
 we used $\Phi$ which is related to the standard dilaton  used in this paper by $\phi=-2\Phi$.}
\bea\label{dilformula}
e^{-2\phi} & = & {1\over 4K^2} \Big( (A + \bar A)K  - (B + \bar B)^2 \Big)
		\Big( (A + \bar A)K - (B - \bar B)^2  \Big) \\
 \chi &=& {i\over 2K} \Big( (A - \bar A)K -B^2 +\bar B^2  \Big) \label{axionform}
\eea
 The component along $M_4$ of the R-R four form potential, $C_K$, is given by
 \bea\label{cfourexp}
C_K= -{i \over 2} {B^2-\bar B^2\over A+\bar A} -{1\over 2}\tilde K
\eea
The expressions for the NS-NS and R-R two-form potentials are given in Appendix \ref{pageeval}.

\subsection{Regular solutions}\label{regsol}
The formulae  presented in the previous section give the local solutions of the BPS equations and equations of motion for arbitrary
choices of the harmonic functions $H$ and $K$, and meromorphic functions $A$ and $B$.  However, further conditions must
 be imposed in order to obtain physically sensible solutions, i.e.
solutions  which are regular and have real-valued physical fields.
In this paper we assume the Riemann surface $\Sigma$ with no handles and a single boundary component,
so that it can be mapped to the upper half plane. The case in which $\Sigma$ has multiple boundary
components has been studied in \cite{Chiodaroli:2009xh}.
Furthermore, the solutions only have $AdS_3 \times S^3\times M_4$ asymptotics and
the meromorphic and harmonic functions have only simple poles.

It was shown in \cite{Chiodaroli:2009yw} that in order to have the proper boundary structure, the real  harmonic functions $H$, $K$, $A+ \bar A$ and $B + \bar B$ all go to zero on the boundary with the same rate. Furthermore, finiteness of the metric factors requires that the harmonic functions $A+ \bar A$, $B + \bar B$ and $K$ have common poles and there is a condition relating the residues of these functions. Singular points in the bulk of $\Sigma$ are forbidden, as are zeroes in the bulk of $\Sigma$ for $A+\bar A$, $K$, and $H$. Finite curvature requires that $B$ and $\partial_z H$ have common zeroes. These conditions restrict the form of the harmonic and meromorphic functions as follows.

The harmonic function $H$ has $n$ distinct simple  poles which lie on the real axis. The $n$th pole is taken to be at infinity.
\be H = i  \sum^{n-1}_{i=1} { c_{H,i} \over z-x_{H, i} } - i {c_{H,n} z} + c.c. \label{multipoleH} \ee
The harmonic function $A$  has $2n-2$ poles which lie on the real axis
\be A = i  \sum^{2n-2}_{i=1} { c_{A,i} \over z-x_{A, i} } +i a \label{multipoleA} \ee
The meromorphic function $B(z)$ is determined in terms of $A$ and $H$ by
\bea\label{bdefine}
B= B_0 { \prod^{n-1}_{i=1} (z-x_{H, i})^2 \over \prod^{2n-2}_{i=1} (z-x_{A, i})} \partial_z H \label{multipoleB}
\eea
The function $K$ is given by
\be K = i \sum^{2n-2}_{i=1} { c_{K,i} \over z - x_{A,i} } -i   \sum^{2n-2}_{i=1} {  c_{K,i} \over \bar z - x_{A,i} } ,\qquad   c_{K,i} = {c^2_{B,i} \over c_{A,i} } \label{multipoleh} \ee
Where the residues $c_{B,i}$ are given by
\be
c_{B,i} ={1\over i} \lim_{z \rightarrow x_{A,i}} (z-x_{A,i}) B (z)
\ee
Note that the dual harmonic function $\tilde K$ which appears in (\ref{cfourexp}) contains another parameter
\bea\label{tildeKdef}
 \tilde K =    \sum^{2n-2}_{i=1} { c_{K,i} \over z - x_{A,i} } +  \sum^{2n-2}_{i=1} { c_{K,i}\over \bar z - x_{A,i} } + 2 k
\eea

\begin{figure}
\centering

\includegraphics[ scale=0.30]{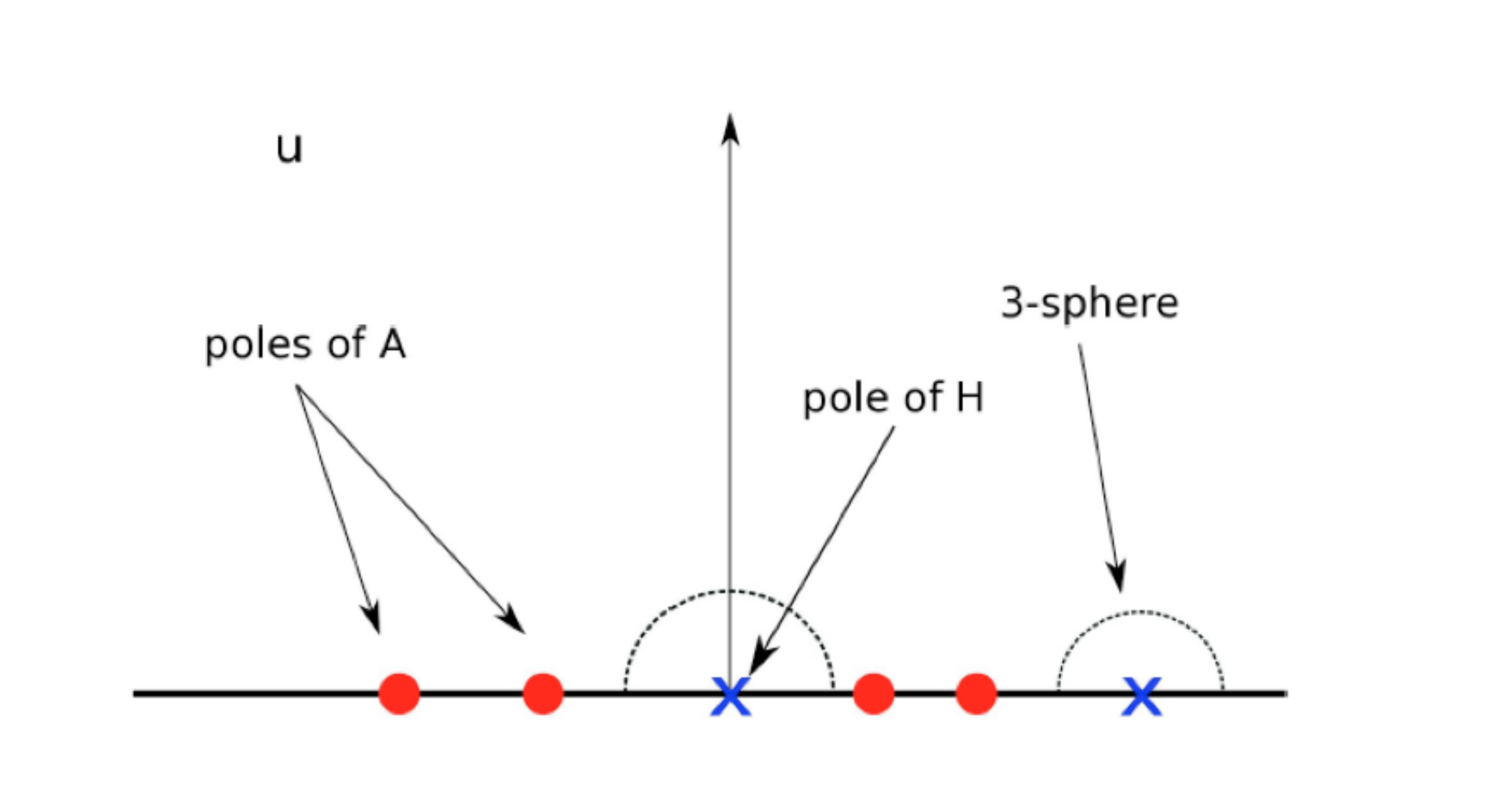}
\caption{Multi-pole solution. The poles of $H$ and $A$ lie on the real axis.}
\label{fig:multi-pole}
\end{figure}

The number of poles of the harmonic function $H$, $n$, also counts the
number of asymptotic $AdS_3$ regions.
Approaching a pole of $H$ in $\Sigma$ corresponds to approaching the
boundary of an asymptotic $AdS_3$ region, while staying away from the
defect. A regular solution with $n$ asymptotic regions depends on the following global parameters:

\begin{enumerate}
\item $n-3$ locations of poles of $H, x_{H,i}$. The 3 is subtracted
 since $SL(2, R)$ reparameterizations of $\Sigma$ can be used
to fix three locations at $x_{H,1} = 0, x_{H,2} = 1$ and $x_{H,n} = \infty$.

\item $n$ positive values of the residues of $H, c_{H,i}$.
\item $2n - 2$ locations of poles of $A, x_{A,i}$.
\item $2n - 2$ positive values of the residues of $A, c_{A,i}$.
\item two real additive constants in $A$ and $\tilde K$, $a$ and $k$.
\item a real multiplicative constant in $B, B_0$.
\end{enumerate}
The total number of parameters, or moduli, for the $n$-pole solution is $6n - 4$. These parameters determine all physical quantities of the solution, the full metric, the charges of the antisymmetric three-form, and the scalars, the axion and the dilaton.

\section{Charges and fields in the asymptotic regions}\label{secthree}
\setcounter{equation}{0}

Since type IIB supergravity is a theory with  Chern-Simons terms, the  Bianchi identities for the antisymmetric tensor fields are modified and
there are three different definitions of charge associated with the branes \cite{Marolf:2000cb}.
These three charges, namely Maxwell, brane source and Page charges,  are reviewed in Appendix \ref{appendixa}.

In this section, we
compute the Page charges   \cite{Marolf:2000cb,Page:1984qv} which are  the conserved charges
associated with the (quantized) number of branes. The dual conformal field theory is characterized by the number of  underlying
five- and one-branes, so the Page charges are the most useful in the
identification of the dual CFT.
 More details of the computation can be found  in Appendix \ref{appendixa}.

\subsection{Five- and one-brane charges}
In type IIB, the Page charges for NS5- and D5-branes are given by \footnote{The fields $\tilde F_3$, $C_4$, $C_2$ are defined in Equation  (\ref{fthreedef}) in Appendix \ref{appendixa}.}
\bea
\label{moo}
Q_{NS5}^{Page}= \int_{S^3} H_3, \quad\quad  Q_{D5}^{Page}\int_{S^3} \Big(\tilde F_3 +\chi H_3\Big)\ 
\eea
The Page charges are localized and conserved and are evaluated in Equation (\ref{pageeval}). The expressions for the five-brane charges in terms of the gauge potentials are given by \footnote{The potentials $b^{(2)}$, $c^{(2)}$ are defined in Equations  (\ref{potdef2b}, \ref{potdef4}) in Appendix \ref{pageeval}.}
\bea
Q_{NS5}^{Page}&=&  4\pi \big(\int_{\cal C} dz \; \partial_z b^{(2)} + c.c \big) \no\\
Q_{D5}^{Page}&=&  4\pi \big(\int_{\cal C} dz \; \partial_z c^{(2)} + c.c \big)
\eea
where ${\cal C}$ is a contour in the Riemann surface $\Sigma$.   
The fibration  of  $S^2$  over ${\cal C}$ produces the $S^3$ in the asymptotic region and this $S^3$ is the integration domain in (\ref{moo}).  For the solutions in this paper, the Riemann surface $\Sigma$ is the half-plane and the contours which produce homology three-spheres are the ones which enclose a pole of the harmonic function $H$.

The Page charge  for the D1-branes is given by
\bea\label{paged1bis}
 Q_{D1} ^{Page } &=&-\int_{M_7}\Big( e^{\phi}* \tilde F_3 -4 C_4\wedge H_3\Big)
 \eea
 and the Page charge associated with the fundamental string is given by
 \bea
 Q_{F1}^{Page } &=& - \int_{M_7} \Big( e^{-\phi} * H_3 -\chi e^\phi *\tilde F_3 +4 C_4\wedge dC_2)
 \eea
 where $M_7$  is a product of $M_4$ and the aforementioned $S^3$.

  The expressions for the one-brane charges  given in (\ref{paged1bis}) are more complicated due to the Hodge dual and the presence of Chern-Simons terms.   The seven manifold is a product of $M_4$ and a homology three-sphere obtained from a contour ${\cal C}$
 in the Riemann surface $\Sigma$ together with the fibered $S^2$, just as in the case of the five-brane charges.

  \medskip

  \noindent The D1-brane charge  is given by
\bea
Q^{Page}_{D1}&=& 4 \pi  \Big\{  \int_{\cal C}  {4 K \over A+\bar A}  {(A+\bar A)K-(B+\bar B)^2\over (A+\bar A)K -(B-\bar B)^2} i(  \partial_z c^{(1)} -\chi \partial_z b^{(1)} ) dz \no\\
&&- 2\int_{\cal C} \Big( { i } {B^2-\bar B^2\over A+\bar A} +\tilde K\Big)  \partial_{ z} b^{(2)} dz  \Big\}+c.c.
\eea
The fundamental string charge is given by
\bea
Q^{Page}_{F1}&=& 4\pi \Big\{ \int_{\cal C} {\Big( (A+\bar A)K-(B+\bar B)^2\Big)^2 \over K(A+\bar A)} i \partial_z b^{(1)} dz +2\Big({i(B^2-\bar B^2)\over A+\bar A} +\tilde K \Big)   \partial_z c^{(2)} dz\no\\
&&- \int_{\cal C}   {4 K \over A+\bar A}  {(A+\bar A)K-(B+\bar B)^2\over (A+\bar A)K -(B-\bar B)^2}    \;   i\chi \Big(   \partial_z c^{(1)} -\chi \partial_z b^{(1)}\Big) dz \Big\}+ c.c.
\eea

The charges associated with each asymptotic $AdS_3$ region can be evaluated by choosing the contour ${\cal C}$ infinitesimally 
close to the associated pole of $H$. In the remaining of the paper we will use only the Page charges, but drop the 
Page label for notational simplicity.

\subsection{Local expansion near a pole}\label{localexp}

In this section we expand the solution near a pole  $z=x_{H,i}$ of the harmonic function  $H$ (\ref{multipoleH}). For notational  simplicity we perform the expansion for a pole located  at $z=0$. Note any pole  $z=x_{H,i}$ can be mapped to 0 by a translation or inversion.
The harmonic and meromorphic functions have the following expansion around $z=0$, where we only keep the leading  terms on which the asymptotic fields and charges depend,
\bea\label{locexpan}
H&=& i \; {c_{-1} \over z } + i \; c_1 z - i\;  {c_{-1} \over \bar z }-  i \;c_1 \bar z + \cdots\no\\
A&=&  i \; a_0 + i \; a_1 z +\cdots \no\\
B&=& i \; b_0 + i \; b_1 z +\cdots \no\\
K&=&  i\; k_1 z - i \; k_1 \bar z  +\cdots  \no\\
\tilde K&=&  2 k_0 + k_1 z +   k_1 \bar z  +\cdots
\eea
These local parameters, $c_{-1},c_{1} ,a_0,a_1 ,b_0 , b_1,k_0$ and $k_1$ are real.  Note that for each pole at $z=x_{H,i}$ one obtains a set of constants.
These constants can be expressed in terms of the moduli of the regular solutions of Section \ref{regsol}.

\subsubsection{The fields and metric}\label{asymval}

The asymptotic values of the various scalar fields can be expressed in terms of the local parameters. One obtains  for dilaton and axion
\bea
e^{-2\phi} &=& {b_0^2 \big( a_1 k_1-b_1^2\big) \over k_1^2}+ o(r), \quad \quad
\chi =  -a_0 + {b_0 b_1\over k_1} +o(r) \label{Asymptotic-chi}
\eea
where $z=r e^{i\theta}, \bar z= r e^{-i\theta}$ and we are expanding $r$ around $r=0$.
The metric factor and the R-R four form potential on $M_4$ are given by
\bea
f_3^4 &=&   {4 b_0 \sqrt{ a_1 k_1-b_1^2} \over a_1}+o(r), \quad \quad
C_K = \Big( {b_0  b_1\over a_1} -{k_0}\Big)+o(r) \label{Asymptotic-Ck}
\eea
The other metric factors have the following expansion around $r=0$,
\bea
\rho^4&=&  {a_1 c_{-1}^2 \sqrt{a_1 k_1 -b_1^2} \over b_0^3} \; {1\over r^4} + o(1/r^3) \no\\
f_1^4  &=& {1\over r^4}  {a_1 b_0 c_{-1}^2 \over (a_1 k_1 -b_1^2 )^{3/2}} + o(1/r^3)  \no\\
f_2^4 &=&  \sin^4(\theta) {a_1 c_{-1}^2  \sqrt{ a_1 k_1-b_1^2}  \over b_0^3} + o(r)
\eea
Defining a new coordinate $x$  for the $AdS_2$ slicing,
\be
r={  2b_0\over \sqrt{ a_1k_1-b_1^2 }}\exp(-x)
\ee
the ten-dimensional metric takes the following asymptotic form as $x\to \infty$,
\be\label{asymmeta}
ds_{10}^2 = l^2\Big(  dx^2 + {1\over 4}\exp({2x  } )\; ds_{AdS_2}^2 \Big)+   l^2  \Big( d\theta^2+ \sin^2\theta ds_{S^2}^2\Big) +   {2b_0^2 \over a_1 c_{-1}} \; l^2 \;ds_{K3}^2+o(e^{-x})
\ee
where $l$ is given by
\bea\label{ldefin}
l=\left({a_1  c_{-1}^2 \sqrt{a_1 k_1 -b_1^2} \over b_0^3} \right)^{1\over 4}
\eea
In the following sections, we will also use the six-dimensional $AdS_3$ radius, defined as
\bea\label{Radsdefin}
R_{AdS_3}= l f_3 =  \left({ 4 c_{-1}^2  (a_1 k_1 -b_1^2) \over b_0^2} \right)^{1\over 4}
\eea

\subsubsection{The charges}

One can read off the central charge $c$ of the dual CFT from the curvature radius of the asymptotic three-dimensional $AdS_3$ metric \cite{Brown:1986nw} in the limit $x\to \infty$,
\bea
ds^2_{AdS_3}= l^2\big(dx^2+ {1\over 4}e^{2x} ds^2_{AdS_2}\big)+ o(e^{-2x}), \quad  c={3 l \over 2 G^{(3)}_N}
\eea
where   $l$ is given by (\ref{ldefin}).
The three dimensional Newton's constant  $G^{(3)}_N$ can also be determined from the ten-dimensional metric  (\ref{asymmeta})
\bea{1\over 16 \pi G^{(3)}_N}= {Vol(M_4) Vol(S^3) \over 2 k_{10}^2}
\eea
Hence, the central charge of the CFT  associated with the asymptotic $AdS_3$ region is given by
\be\label{centralch}
c= {12 \pi\over k_{10}^2} l Vol(M_4) Vol(S^3) =  {96 \pi^3\over k_{10}^2} {   c_{-1}^2 \big(a_1 k_1 -b_1^2 \big)\over b_0^2}
\ee
The D5- and NS5-brane charges given in Section \ref{page5ch}  can be expressed as
\bea\label{Page-NS}
Q_{D5}^{} &=& 4\pi^2  c_{-1}  {a_1 b_0 - a_0 b_1 \over b_0^2}  \no\\
Q_{NS5}^{} &=& 4 \pi^2 c_{-1} {b_1 \over b_0^2}
\eea
Using the  formulae for the Page charges given in Section \ref{paged1ch} and \ref{pagef1ch} one obtains
\bea
Q_{D1}^{} &=& - 16 \pi^2    c_{-1}   { \Big( b_1 k_0 -b_0 k_1\Big) \over b_0^2}\no\\
Q_{F1}^{} &=& -16 \pi^2   c_{-1} { b_0^2 b_1 + a_0 b_1 k_0 - a_1 b_0 k_0 -a_0 b_0 k_1 \over b_0^2}
\label{PageNS} \eea
Using the expressions for the charges (\ref{Page-NS}) and (\ref{PageNS}), the central charge (\ref{centralch})
 can be expressed as follows in terms of the Page charges,
\bea\label{centcha}
c= {6 \over 4 \pi  k_{10}^2 }  \Big(Q_{D1}Q_{D5}+ Q_{F1}Q_{NS5} \Big)
\eea

\subsection{Duality transformations}

Here we look at the transformation of the charges and fields under T-duality of the four internal directions and S-duality. T-duality exchanges $A(z)$ with the holomorphic part of $K$,
\be
T:  \quad  A(z) \leftrightarrow \text{Hol}(K)(z)
\ee
Under this transformation the R-R scalar fields and charges transform as
\bea
T: \quad \quad \chi \leftrightarrow   C_K,\quad \quad   f_3^4 \leftrightarrow  4  e^{-\phi},\quad \quad Q_{D1} \leftrightarrow  {4  } Q_{D5}
\eea
The S-duality transformations act as follows on the meromorphic functions
\be
S:\quad   A \rightarrow {1 \over A}, \qquad B \rightarrow i {B \over A}, \qquad  K \rightarrow K +i  {B^2 \over A} - i {\bar B^2 \over \bar A}
\ee
Under  S-duality the fields transform in the following way
\bea
  S:\quad \quad C_K\to C_K, \quad f_3^4 \to f_3^4, \quad \tau\to -{1\over \tau}
\eea
where the complexified scalar is
\bea
\tau =\chi+ i e^{-\phi}
\eea
The Page charges transform as
\bea
S:\;\;\left(
\begin{array}{c}
   Q_{D1}\\
    Q_{F1}\\
\end{array}
\right) \to \left(
\begin{array}{cc}
   0&1 \\
   -1& 0\\
\end{array}
\right)  \left(
\begin{array}{c}
   Q_{D1} \\
    Q_{F1}\\
\end{array}
\right), \quad\left(
\begin{array}{c}
   Q_{D5}\\
    Q_{NS5}\\
\end{array}
\right) \to \left(
\begin{array}{cc}
   0&1 \\
   -1& 0\\
\end{array}
\right)  \left(
\begin{array}{c}
   Q_{D5} \\
    Q_{NS5}\\
\end{array}
\right)
\eea

\subsection{Holographic junctions}
The  special case in which the solutions presented in Section \ref{regsol}
have $n=2$ asymptotic regions corresponds to the BPS Janus solution which was discussed in detail in \cite{Chiodaroli:2009yw}. 

In this section we generalize the argument given in \cite{Chiodaroli:2009yw}
to show that the holographic interpretation of the solution with $n>2$ regions is a junction of $n$ 1+1-dimensional CFTs,
all of which live on a spatial half line and are  joined at a 0+1-dimensional point.
As shown in Section \ref{asymval}, the expansion near the pole of $H$   produces the metric (\ref {asymmeta}). To express the holographic boundary it is useful to change coordinates to $u= e^{-x}$,
\bea\label{asymmeta2}
ds_{10}^2 &=&{l^2 \over u^2} \Big(  du^2  +  \frac{1}{4} ds_{AdS_2}^2  +   u^2  (d\theta^2+ \sin^2\theta ds_{S^2}^2 )+  {2b_0^2 \over a_1 c_{-1}}  u^2 \;ds_{M_4}^2\Big) +o(u^2)
\eea
 A  boundary  component of the bulk space away from the defect is reached by taking $x\to \infty$ which translates into $u\to 0$. 
In this limit,  the overall  conformal factor diverges.  The boundary geometry is then obtained by removing the divergent conformal factor. 
Note that the $AdS_2$ part of the metric is the only term which does not vanish in this limit.
The standard Poincare patch $AdS_2$ metric  is given by
\bea\label{poincare}
ds_{AdS_2,P}^2 ={-dt^2 +d\xi^2 \over \xi^2}
\eea
After removing the divergent conformal factor $1/u^2$, the boundary geometry  is $R\times R^+$  spanned by the time coordinate $t\in R$ and $\xi \in R^+$. 
For each pole of $H$ we obtain a half-plane as the  associated   boundary component.
Since the poles of $H$ are all separated along the boundary of $\Sigma$, it may seem that the $n$ boundary components are disconnected.
This is however not true since there is an additional boundary component, namely the boundary of $AdS_2$. In the limit $\xi\to 0$ the ten-dimensional metric
behaves as follows
\bea
ds_{10}^2 & \sim&  {1\over \xi^2} \Big( \xi^2 f_2 ^2  ds_{S^2}^2+ \xi^2 f_3^2 ds_{M_4}^2 + \xi^2 \rho^2 dz\bar d\bar z +f_1^2 (d\xi^2  -dt^2)\Big)
\eea
Stripping away the divergent conformal factor shows that the holographic  boundary is simply given by a point $\xi=0$ times $t$.  Since the metric factor of $\Sigma$ vanishes in this limit the distance between different poles goes to zero. This implies that the $n$ half spaces are glued together at a point $\xi=0$. Consequently   the spatial part of the  holographic boundary is  a junction, where $n$ half lines are joined at a point. If one uses global coordinates for $AdS_2$ instead of the Poincare coordinates (\ref{poincare}),
\be
ds_{AdS_2,G} ^2={1\over \sin ^2\sigma} \big(-dt^2+d \sigma^2\big)
\ee
 the conformal boundary has the structure of $n$ finite  intervals $\sigma=[0,\pi]$ which are  joined at  two end points $\sigma=0,\pi$.  For the case of $n=2$, i.e. the Janus solution, the boundary structure can be made more precise by introducing Fefferman-Graham like coordinates \cite{Papadimitriou:2004rz}. The construction of Fefferman-Graham coordinates for  case $n>3$ is mathematically more involved and will not be pursued here.

\section{Supersymmetry and attractor mechanism}\label{attracmech}
\setcounter{equation}{0}
In this section we analyze the conditions for junctions of strings to preserve a fraction of the original supersymmetries.
First we review the case of $(p,q)$ string junctions  in flat space for  ten-dimensional type IIB string theory.
This analysis is then generalized for string junctions  in six-dimensional ${ \cal N}=(2,0)$ supergravity.
We find supersymmetric junctions with an arbitrary number of stings and conjecture that our solutions describe their near-horizon limit.
Furthermore we discuss the attractor mechanism in six dimensions.

\subsection{String junction of $(p,q)$-strings}\label{pgstring}

In this section we review the conditions which need to be satisfied so that  a junction of $(p,q)$ strings in type IIB preserves
a quarter of the original supersymmetry \cite{Sen:1997xi}. The type IIB supersymmetric algebra is given by
\bea
\{ Q^i_\alpha, Q^j_{\beta}\} = (P_+ C \Gamma^\mu)_{\alpha\beta}  \Big( \delta^{ij} P_\mu +\sigma_3^{ij} Z_\mu+ \sigma_1^{ij} \tilde Z_\mu\Big)
\eea
Here $Q^i_{\alpha},i=1,2$ are two Majorana-Weyl supercharges.
We have only kept the central charges $Z_\mu, \tilde Z_\mu$ associated with fundamental strings and $D1$-branes respectively.
For static configurations, $P_0$ gives the tension of the string. A single static $(p,q)$ string has tension
\bea
P_0=T = \sqrt{ Z^2+ \tilde Z^2} =  \sqrt{ (p+\chi q)^2+ e^{-2\phi}q^2 }
\eea
It was shown in \cite{Sen:1997xi} that in order for a junction to be 1/4  supersymmetric,
all the strings  lie in a plane - say the 1-2 plane - and the orientation of the i-th string must be determined by the central charges
\bea
Z_1+i Z_2 &=&  | p_i+q_i \chi |  e^{i \theta_{i} } , \quad   \tilde Z_1+i \tilde Z_2 =  | q_i e^{-\phi} |  e^{i \theta_{i} }
\eea
where  the angle $\theta_{i}$ is determined by
\bea
p_i+ q_i\tau = |p_i + q_i \tau| e^{i \theta_i}
\eea
Three $(p,q)$ strings which meet at a point  are oriented in the plane such  that the tensions of the strings in the three string junction balance.
In particular we have charge conservation
\bea
\sum_i{q_i}=0, \quad \sum_i p_i =0
\eea
and tension balance
\bea
\sum e^{i\theta_i} T_i
\eea

\begin{figure}
\centering
\includegraphics[ scale=0.40]{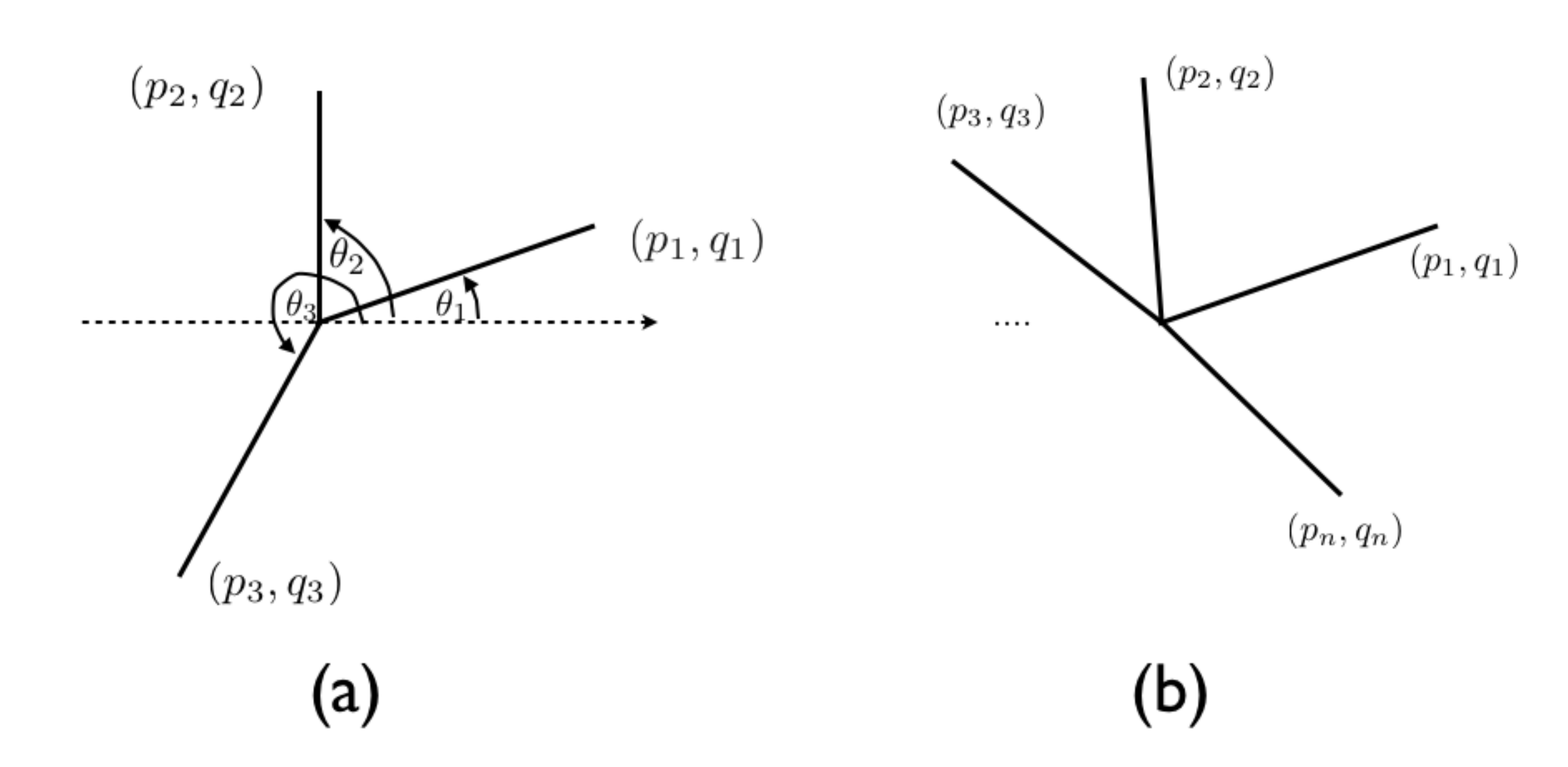}
\caption{(a) Three string junction (b) multi-string junction.}
\label{fig1}
\end{figure}

\subsection{String junctions in six dimensions} \label{secfiveb}

Type IIB supergravity compactified on $K_3$ gives the six-dimensional ${\cal N}=(2,0)$ supergravity  \cite{Romans:1986er}
which has sixteen real supersymmetry generators. The scalars live in the coset
\bea\label{coseta}
 {\cal M}={SO(5,21) \over SO(5) \times SO(21)}
\eea
which can be parameterized by a coset vielbein
\be (V^{i}_A, V^{r}_A) \label{exp-Viel} \ee
with  $i={1,2,\cdots 5}$ and  $r=1,2,\cdots 21$. The field strengths $H^{i}$ and $H^{r}$ are self dual and anti self dual respectively
  \bea
H^{i} &=&\sum_A  V^{i}_{\;A} G^{A} , \quad \quad  H^{i}= *H^{i} , \quad \quad i=1,2\cdots ,5\no\\
H^{r} &=&\sum_A V^{r}_{\;A} G^A, \quad \quad H^{r}=-*H^{r} , \quad \quad r=1,2\cdots ,21
\eea
The third rank antisymmetric tensor fields $G^{A}$ with $A=1,2, \cdots,26$ satisfy  simple Bianchi identities and the charges
\bea\label{qcharga}
Q^A =\int_{S^3} G^A, \quad A=1,2, \cdots, 26
\eea
are quantized and lie in a lattice $\Gamma^{5,21}$.  The supergravity theory has a
multitude of BPS dyonic string solutions  which carry  charges in the charge lattice $\Gamma^{5,21}$.
The ten-dimensional interpretation of these strings is that they come from
 D1-branes and fundamental strings  in six dimensions, D5- and NS5-branes wrapping the $M_4$  and D3-branes wrapping two-cycles on $M_4$.
 The six-dimensional supersymmetry algebra is given by
 \bea\label{susyalgebra}
 \{  Q_\alpha^a, Q_\beta^b\} = (P_+ C_{(6)} \gamma_{(6)}^\mu)_{\alpha\beta}\Big( C_{(4)}^{ab} P_\mu + (C_{(4)} \Gamma_i)^{ab} Z_\mu^i \Big)
 \eea
Where $C_{(6)}$ and $C_{(4)}$ denote  the six-dimensional and four-dimensional charge conjugation matrices.
The supercharges $ Q_\alpha^a$ are six-dimensional simplectic  Majorana spinors, where $a=1,\cdots 4$ is an $SO(5)$
spinor index. $Z_i, i=1,2,\cdots 5$ are central charges which transform as a vector of $SO(5)$.
The central charge which appears in the supersymmetry algebra is related to the  charges (\ref{qcharga}) via the vielbein (\ref{exp-Viel}).
A static string  has a worldvolume which is spanned by the time direction and a unit norm vector $\hat{n}$ in $R^5$ which
indicates the spatial orientation of the worldvolume. For this  static string the time component of
 the central charges $Z^i_0$ vanishes and the spatial components $Z^i_m, m=1,2,\cdots,5$ are given by
\be
Z^i_m =  (\hat{n})_m\sum_A \int_{S^3_{\hat{n}}} V^i_A G^A
\ee
Where $S^3_{\hat{n}}$ is the three-sphere  at infinity in the four directions transverse to $\hat{n}$.
A single string preserves half the supersymmetry if the following condition
\bea\label{susyproj}
\Big(P_0+ \gamma_0 \gamma^m \Gamma_i Z^i_m \Big) \epsilon=0
\eea
has eight linearly independent solutions.
The condition above translates into the following condition for the tension $P_0$,
\bea\label{tension}
 T=P_0= \sqrt{\sum_{i} \sum_{m}(Z^i_m)^2}
\eea

For  each string we can choose an unit-norm vector $\hat{n}_{(k)}$ in $R^5$.
In order for the string junction to preserve a quarter of the supersymmetry, we need the central charge of  the $k$-th string to to have
only spatial components given by
\bea\label{centrach}
\left(Z^{(k)}\right)_m^i = (\hat{n}_{(k)})_ m \;    \left(R\cdot  \hat{n}_{(k)}\right)^i \;   |T_{(k)}|
\eea
$R $ is  a $SO(5)$ rotation matrix, which is the same for all strings.
It is possible to perform a global R-symmetry rotation which sets $R=1$.
The condition   for two strings with directions  $\hat{n}_{(k_1)}$ and $\hat{n}_{(k_2)}$ to possess compatible supersymmetries is given by
\bea\label{spinorconb}
\Big([ \gamma_m \hat{n}_{(k_1)}^m,\gamma_n\hat{n}_{(k_2)}^n]+ [\Gamma^i \hat{n}_{(k_1)}^i,\Gamma^j\hat{n}_{(k_2)}^j]\Big) \epsilon=0
\eea
If we choose without loss of generality that the unit vectors $\hat{n}_{k_1}$ and  $\hat{n}_{k_2}$ lie in the 1-2 plane then (\ref{spinorconb}) is equivalent to
\be\label{susyconb}
\sin(\theta_{12}) \Big( 1+ \gamma_{12} \Gamma^{12}\Big)\epsilon=0
\ee
Where $\theta_{12}$ is the angle between $\hat{n}_{k_1}$ and $\hat{n}_{k_2}$.
The condition (\ref{susyconb}) is the same for all strings in the 1-2 plane
whose spatial and central charge orientations are both given by $\hat n_k$.
Hence, an arbitrary number of such strings meeting at a junction  preserve the  same four supersymmetries.
Charge conservation then implies that   the  tension balance condition holds
\bea
\sum_k T_{(k)} \vec{n}_{(k)}=0
\eea
A single half-BPS dyonic string solution in six-dimensional supergravity which carries the
D1-, F1-, NS5- and D5-charges the near-horizon limit produces an $AdS_3 \times S^3$ maximally supersymmetric vacuum.
The fact that we can obtain supersymmetric junctions with an arbitrary number of strings leads to
the conjecture that \emph{our half-BPS supergravity solution give the near-horizon limit of a junction of self-dual BPS strings in six flat dimensions}.
Note that the decoupling limit is expected to enhance the supersymmetries from four to eight, which is the number of preserved
supersymmetries of the  half-BPS interface solutions.

\subsection{The $SO(2,2)$ truncation and the attractor mechanism}

The supergravity solution presented in Section \ref{sectwo} has four nontrivial scalars and four non vanishing  antisymmetric tensor fields.
In particular all internal moduli of the $M_4$ surface and all antisymmetric tensor fields associated with the two-cycles of $M_4$
are trivial.

Consequently the solution can be expressed as a consistent truncation of the  full ${ \cal N}=(2,0)$ supergravity.
The full  scalar coset (\ref{coseta}) can be truncated to  scalars living in the coset $SO(2,2)/ SO(2)\times SO(2)$.
The $\Gamma^{5,21}$ charge lattice   is truncated   to a
$\Gamma^{2,2}$ charge lattice.   More details on the reduction can be found in Appendix \ref{sixdred}.

We know that the string tension must be invariant with respect to the global $SO(2,2)$ transformations.  The reduced central charge $Z_i$ is given by
\be Z^{i} = V^i_{\;A} Q^A, \quad \quad i=1,2  \ee
 $Z^i$ transforms only under the $SO(2)$ subgroup of the $SO(5)$
R-symmetry and can be identified with the central charge appearing in the supersymmetry algebra (\ref{susyalgebra}).
The $SO(2)$ invariant
\be T= |Z| =  \sqrt{ (V^1_{\;A} Q^A)^2 + (V^2_{\; A} Q^A)^2  }  \ee
 reproduces the correct tension  formulae   for   the D1/D5 and $(p,q)$ systems. Plugging in our expressions for vielbein and charges we obtain,
\be |Z| = {e^{\phi \over 2} \over f_3^2} \sqrt{ \Big( e^{- \phi} Q_{D1} + f^4_3 \tilde Q_{D5} - 4 e^{- \phi} C_K  Q_{NS5} \Big)^2 +
 \Big( Q_{F1} + \chi Q_{D1} + e^{-\phi} f^4_3  Q_{NS5} + 4 C_K \tilde Q_{D5}\Big)^2 } \qquad  \ee
with $\tilde Q_{D5} =Q_{D5} - \chi Q_{NS5}$. All the charges in the expression above are Page charges. Note that this expression is  T and S-duality invariant and reproduces
 the expressions in the literature for the particular
cases of the $D1/D5$ and $(p,q)$ systems.

The six-dimensional perspective can also help understand which scalars have their asymptotic values fixed by the Page charges.
In particular, the six-dimensional equation of motion for the scalars is
\be D^\mu P^{ir}_\mu = {\sqrt{2}\over 3} H^i_{\mu \nu \rho} H^{r \mu \nu \rho} \ee
where $P^{ir}_\mu$ is the coset one-form field strength and $i$,$r$ are indices corresponding to the two different $SO(2)$ groups.
If we consider a six-dimensional dyonic string solution, the scalars must be constant in the near-horizon region.
This leads to the condition,
\be D^\mu P^{ir}_\mu = 0 \; \rightarrow \; H^i_{\mu \nu \rho} H^{r  \mu \nu \rho} = 0  \; \rightarrow \; V^i_{\ A} V^r_{\ B} Q^A Q^B = 0, \quad i=1,2, \;\; r=\dot{1},\dot{2} \ee
There are two possible solutions:
\be  V^i_{\ A}   Q^A = 0 \quad i=1,2 \qquad \text{or } \quad   V^r_{\ B} Q^B = 0 \quad r=1,2   \ee
The first solution corresponds to a non-BPS extremal attractor and is not interesting to us. The second solution can be rewritten in terms of the asymptotic fields and charges   as
\bea 4 C_K {e^{-{\phi\over 2}} \over f^2_3} + \chi e^{{\phi\over 2}}f^2_3&=& - {Q_{F1} \over Q_{D1}}e^{{\phi\over 2}}f^2_3 + \big( f^8_3 + 16 C^2_K \big) {e^{-{\phi\over 2}} \over f^2_3} {Q_{NS5} \over Q_{D1}} \no \\
{e^{-{\phi\over 2}} \over f^2_3} &=& {Q_{D5} \over Q_{D1}}e^{{\phi\over 2}}f^2_3 + \big( 4 C_K {e^{-{\phi\over 2}} \over f^2_3} - \chi e^{{\phi\over 2}}f^2_3\big) {Q_{NS5} \over Q_{D1}}  \label{attractor}\eea

We see that in general the particular combination of scalars fixed by the attractor mechanism depends on the charges.

\subsection{Expressing the moduli in terms of physical parameters}

In Section \ref{regsol} it was shown that the solution with $n$ poles of $H$ depends on  $6n-4$ moduli. The analysis of Section \ref{localexp} showed that in general each asymptotic region associated with a   pole of $H$ carries four charges, namely   $D1,F1$-branes as well as wrapped $D5$- and $NS5$-branes.
Taking charge conservation into account, this means that the solution carries $4n-4$ independent charges.
From the attractor mechanism discussed in the previous section, it follows that at each pole there are two combinations of
the four scalars which are   attracted and fixed by the charges  and two combinations which are not attracted and can take any value.
Consequently  for each pole there are two asymptotic values for the unattracted  scalars, which are dual to sources for marginal
 deformations of the associated CFT. In summary there are $2n$ physical parameters in addition to the $4n-4$ charges,
 in agreement with the number of moduli of the solution, $6n-4$.

The requirement that the junction only has Ramond-Ramond charges leads to simple conditions involving poles and residues of the relevant
harmonic and meromorphic functions. However, it is a mathematical challenge to express all the global moduli of the solution in terms of the physical parameters.
This problem can be solved only numerically or perturbatively in special cases.

\section{The three-string junction}\label{Three-Jct}
\setcounter{equation}{0}

In this section we present the special case in which the regular solutions given in Section \ref {regsol} have three asymptotic regions.
The Riemann surface $\Sigma$ is then given by the upper half-plane with three poles of $H$ located at $z=0,1,\infty$.
The other relevant functions $A, B $ and $K$ have four poles on the real axis, in positions $p_1, p_2, p_3, p_4$.
These solutions are dual to  junctions of three two-dimensional CFTs defined on half-spaces
joined at a $0+1$-dimensional interface, as illustrated in Figure \ref{fig3}.

We have the following expressions for the harmonic function $H$,
\bea H =  { 2 y c_{H,0}  \over x^2 +y^2 } +   {2 y  c_{H,1}  \over (x-1)^2 +y^2 } + 2 y c_{H,\infty}  \label{3poleH} \eea
where $z=x+iy$.
Similarly, Equation (\ref{multipoleA}) reduces to
\be A + \bar A  =  2 y \sum^{4}_{i=1} { c_{A,i} \over (x-p_{ i})^2 + y^2   }, \qquad -i (A - \bar A)  =  2  \sum^{4}_{i=1} { c_{A,i} (x-p_i) \over (x-p_{ i})^2 + y^2   } + 2 a  \label{3poleA}  \ee
The other functions have similar expressions. As explained in Section \ref{regsol}, the residues of $B$ and $K$ are not independent. 
Given the definition of $B$ in  (\ref{bdefine}), we can express the residues $c_{B,i}$ in terms of the residues of $H$ and the positions of the poles of $A$,
\be c_{B,i} = - B_0 {c_{H,\infty} p^2_i (p_i-1)^2 + c_{H,1} p^2_i + c_{H,0} (p_i-1)^2 \over \prod_{j \neq i} (p_i-p_j) }, \qquad i=1 \dots 4
\label{resB} \ee

The harmonic and meromorphic functions depend on a total of fourteen parameters, namely the positions of the poles of $A$, $p_1 \dots p_4$,
the residues of $H$, $c_{H,0}, c_{H,1}, c_{H,\infty}$, the residues of $A$, $c_{A,1} \dots c_{A,4}$, two additive constants in the definition of
$A$ and $\tilde K$, $a$, $k$, and a multiplicative constant in the definition of $B$, $B_0$.

The main advantage of parameterizing our solutions with poles and residues of the various functions is that
global regularity is guaranteed provided that all the residues are positive and that the poles are on the real axis.
On the other hand, it is difficult to obtain examples with particular asymptotic values of physical fields and charges
and, consequently, to study the dual CFT.
It would be desirable to express the solutions using the physical fields and charges as independent parameters.
 However, the relation between global and physical parameters is non-linear and the change of parameters
can be performed analytically only in particular cases, such as the degeneration limits studied in Section \ref{Sec-degeneration}.

\subsection{Pure Ramond-Ramond junctions}

In order to have junctions with simple dual conformal field theories, it is helpful to set to zero the Neveu-Schwarz
charges in all asymptotic regions.

In this case, each of the three asymptotic regions will be dual to a  conformal field theory  given by the IR fixed point of
a two-dimensional $\mathcal{N}=(4,4)$ theory defined on the worldvolume of a $D1$-$D5$ systems.

The requirement that the $NS5$-charge is zero is equivalent to demanding that the expansion coefficient $b_1$ vanishes in all asymptotic regions.
Using (\ref{resB}) we can explicitly evaluate the coefficient $b_1$ for the region at $z=0$,
\be b_1\big|_{z=0} = c_{H,0} {\sum_i {1 \over p_i} - 2 \over \prod_i p_i} \ee
The vanishing of the NS5 charge then leads to
\be \sum_i {1 \over p_i} = 2 \ee
Similarly, the same condition for the pole at $z= \infty$ is
\be \sum_i p_i = 2 \ee
The vanishing of the $NS5$-charge in the third region is automatic due to charge conservation.
Furthermore, we need to set the fundamental string charge to zero as well. This leads to the conditions
\be a_1 k_0 + a_0 k_1 = 0 \ee
in two asymptotic regions. More explicitly, the above conditions can be rewritten as follows
\bea \Big( \sum_i {c_{A,i} \over p^2_i} \Big) \Big( \sum_i {c^2_{B,i}  \over c_{A,i} p_i} - k \Big) + \Big( \sum_i {c^2_{B,i}  \over c_{A,i} p^2_i} \Big) \Big( \sum_i {c_{A,i} \over p_i} - a \Big) &=& 0 \\
\big( \sum_i {c_{A,i} } \big) k + \big( \sum_i {c^2_{B,i} \over c_{A,i} } \big) a &=& 0
  \eea
These equations are linear in $a$ and $k$ and admit a simple solution.
Finally, we note that the resulting six-dimensional dilaton $r_a$ has a simple expression in terms of the local parameter $b_0$,
\be r^4_a= e^{- \phi} f^4_3 \Big|_{z=a} =4 b_0^2 \Big|_{z=a}, \qquad a={0,\infty} \label{sixddila}  \ee

\subsection{Degeneration limits}\label{Sec-degeneration}

A punctured Riemann surface admits interesting degeneration limits in which some of the punctures become very close to each other.
In string perturbation theory, this degeneration limit corresponds to an amplitude factorizing into lower order amplitudes.
Another recent example arises in the context of Gaiotto dualities \cite{Gaiotto:2009we}, where in the degeneration limit a quiver gauge theory becomes
weekly coupled.

\begin{center}
 \begin{figure}
\includegraphics[width=6.0in]{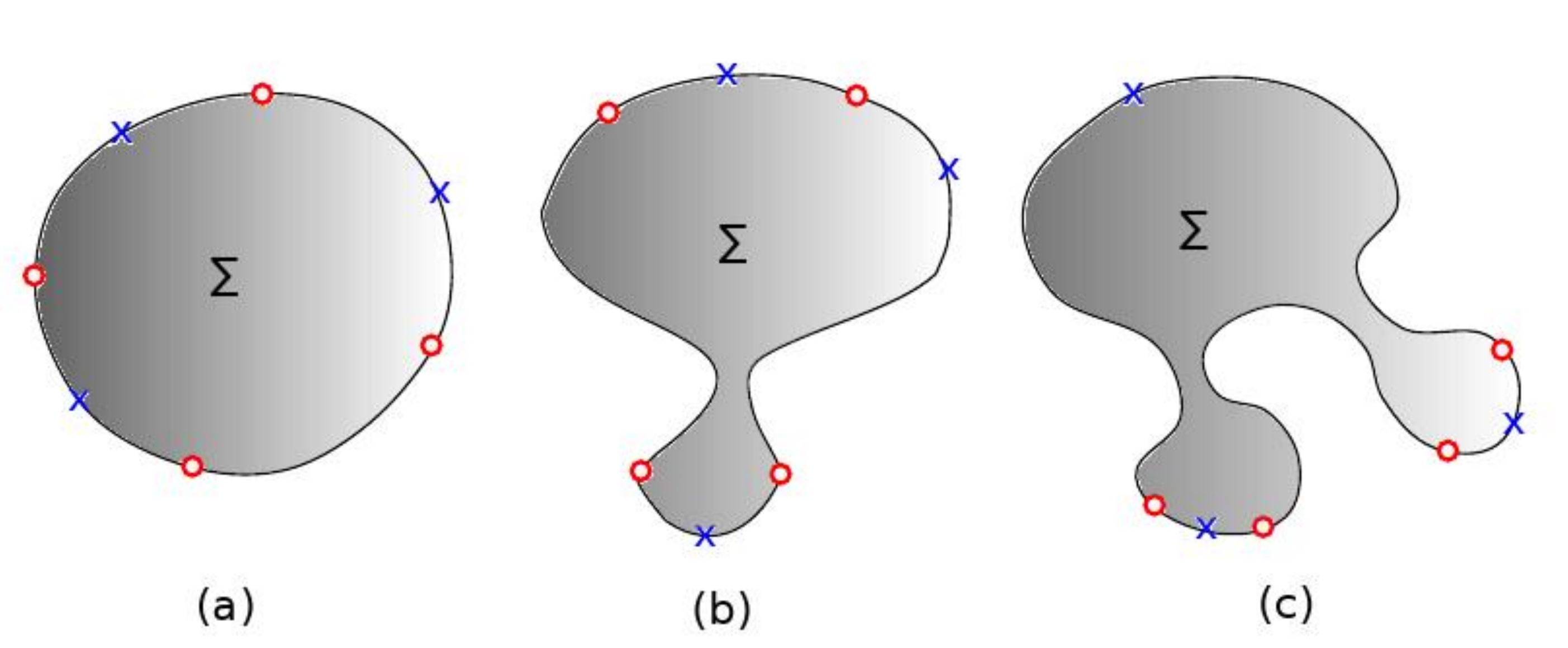}
\caption{Degeneration limit of a three-junction \label{fig-deg}. The junction is non-degenerate in (a). (b) depicts the probe limit while
(c) is the degeneration limit studied in Section \ref{Sec-doubledeg}. }
\end{figure}
\end{center}

More specifically, we can consider the limit of our solutions in which two poles of $A$ collide with a pole of $H$, i.e. approach
one of the asymptotic $AdS_3 \times S^3 \times M_4$ regions. If the corresponding residues
of $A$ and $H$ go to zero in the appropriate way, we obtain a solution where the charges in one of the regions
are very small compared to the ones in the other regions. In this limit, the three-pole solution reduces to
a probe brane in a background given by the two-pole Janus solution studied in \cite{Chiodaroli:2009yw} and \cite{Chiodaroli:2010ur}. This limit will be studied
in the next subsection and is depicted in Figure \ref{fig-deg} (b).

Another example is the (doubly-degenerate) limit in which two poles of $A$ become very close to the asymptotic region at $z=0$ and
the  other two poles of $A$ become very close to
$z = \infty$. This limit is illustrated graphically in Figure \ref{fig-deg} (c). Part of the reason why the
limit is interesting is that the Neveu-Schwarz charges of the solution can be set to zero giving a solution
with a clear CFT interpretation; a solution where only one pole of $A$ collides with an asymptotic region
would have non-vanishing $NS5$-charge. Moreover, we will see that, in contrast to the probe limit,
this degeneration limit allows for finite charges and vanishing jump in the unattracted axion.

In Section \ref{holoent}, the entanglement entropy is calculated in these two degeneration limits and expressed in terms of the physical
data.

\subsubsection{The probe limit}\label{probe}

Without any loss of generality, we will assume the probe to be located at $z=1$. In the following analysis we will
denote the charges in the asymptotic region at $z=0$ by $Q_{D1}, Q_{D5}$ and the probe charges by $ \epsilon q_{D1}, \epsilon q_{D5}$
with  $\epsilon \ll 1$.
The charges in the third region at $z=\infty$ are determined by charge conservation.
We will take the meromorphic function $A$ in the form
\be A(z)= i \left( {c_{A,1} \over z - p_1} + {c_{A,2} \over z - p_2} + {\sqrt{\gamma} / \lambda \epsilon  \over z - 1 - \alpha \epsilon}
+ {\sqrt{\gamma}  \lambda \epsilon  \over z - 1 + \beta \epsilon}  + a \right)    \ee
and the harmonic function $H$ in the form
\be H(z,\bar z) = i \left( {c_{H,0} \over z} + {\omega^2 \epsilon^2 \over z-1} -  z c_{H,\infty}  \right) + c.c. \ee
With these definitions, the probe parameters $\alpha, \beta, \gamma, \lambda$ and $\omega$ are all of order one.
Note that the residues of the functions $A,B$ and $K$ which correspond to the poles colliding with the  $z=1$ asymptotic regions
all vanish as
\be c_{B,i},c_{K,i} = O( \epsilon), \qquad i=3,4 \ee
so that we can recover the two-pole solution for $\epsilon=0$.

We start by studying the solution to order $\epsilon^0$, i.e. expressing the moduli in terms of the background physical parameters. For simplicity, we can
set to zero the additive constants in the definition of $A$ and $K$,
\be a=k=0 \ee
Note that with this choice the fundamental string charge of the background geometry is automatically zero.
As seen before, it is particularly convenient to study a pure
Ramond-Ramond solution, and the condition for the vanishing of the $NS5$-brane charge reduces to a simple relation between the positions of
the first two poles of $A$,
\be p_1=-p_2= p \ee
The other global parameters of the solution can be expressed in terms of the physical fields of the background geometry.
The residues of $A$ are given by
\be  c_{A,1}+c_{A,2} =- {Q_{D5} B_0 \over 4 \pi^2}, \qquad  c_{A,1} - c_{A,2} =  p \Delta \chi_0
\label{A-bg}  \ee
The residues of $H$ are
\be c_{H,0}=-p^2{ r_0^2 \over 2 B_0}, \qquad  c_{H,\infty}= - { r_\infty^2  \over 2 B_0 }  \ee
and the constant $B_0$ has the expression
\be B_0 =  - 4 \pi^2 p \sqrt{{(r_0^2+r_\infty^2)^2 \over Q_{D1} Q_{D5} } + {\Delta \chi_0^2 \over Q^2_{D5} }} \ee
In the above expressions, $r_0$ and $r_\infty$ are defined as in (\ref{sixddila}).

Similarly, $\Delta \chi_0=\chi_0- \chi_\infty=\chi_0$ denotes the jump in the axion in the background geometry.
We can assume without any loss of generality that $Q_{D1},Q_{D5} > 0$. This leads to a negative value for the constant $B_0$. We will also assume
that the probe charges have the same sign, since we know that a brane with charges of opposite sign breaks all supersymmetries.

The position of the first two poles of $A$, denoted by $p$, has not been fixed yet.
Note that the scaling $p \rightarrow \text{const} \   p$
rescales all the residues, but leaves the physical fields unchanged. This symmetry correspond to a rescaling of the complex coordinate
$z$ and could be used to set $p=1$ if the probe brane was not present.

So far the probe brane has not been included in the analysis. Next, we consider the order $\epsilon$ terms in our solution.
As in the analysis of the background geometry, we will consider a probe brane with vanishing Neveu-Schwarz charges.
The vanishing of the $NS5$-brane charge leads to
\be \alpha = \beta  \label{expr-beta} \ee
In other words, to have vanishing $NS5$-charge we need the extra two poles of $A$ to approach the $z=1$ region symmetrically.
Moreover, the vanishing of the fundamental string charge can be used to find an expression for the parameter $\gamma$,
\be \gamma= {c_{A,1} c_{A,2} p^2 \over \alpha^2 (p^2-1)^2 } \Big( { (c_{H,0} + c_{H, \infty}  ) \alpha^2 + \omega^2 \over c_{H,0} + p^2 c_{H, \infty}  } \Big)^2
{(p+1) c_{A,1}- (p-1) c_{A,2} \over (p-1) c_{A,1} - (p+1) c_{A,2} } \label{expr-gamma} \ee
Global regularity demands that all the residues of $A$ are real and positive. To avoid producing two imaginary residues,
we need $\gamma$ to be positive. It is immediate to see that this requirement can be satisfied only if
\be c_{A,1} \geq {p+1 \over p-1} c_{A,2}, \; \; \; p>1 \qquad \text{or} \qquad c_{A,2} \geq {p+1 \over p-1} c_{A,1}, \; \; \; p>1 \ee
These conditions are quite interesting: in order to have a regular solution, the residues of $A$ cannot be equal to each other.
In other words, a symmetric pure Ramond-Ramond junction must be singular in the probe limit.
In particular, we cannot have a regular solution for a back reacted Ramond-Ramond brane in the $AdS_3$ Ramond-Ramond vacuum.
Moreover, because of (\ref{A-bg}), a regular solution needs to have a jump in the unattracted axionic fields at least in the probe limit.

The next step is to set the value of the six-dimensional dilaton in the probe region, which will be denoted with $r_1$.
This can be done by fixing the parameter $\omega$ to
\be \omega^2 = - \alpha^2 { p^2-1 \over 2 B_0} r_1^2 \label{expr-omega} \ee
Moreover, one can find an expression for the ratio of charges of the probe brane,
\be {q_{D1} \over q_{D5}} = {Q_{D1} \over Q_{D5}} {p \Delta \chi_0 - \sqrt{ (r_0^2 + r_\infty^2)^2 {\textstyle Q_{D5} \over \textstyle Q_{D1}}+\Delta \chi_0^2  }
\over p \Delta \chi_0 + \sqrt{ (r_0^2 + r_\infty^2)^2 { \textstyle Q_{D5} \over \textstyle Q_{D1}}+\Delta \chi_0^2  } } \ee
Since the parameters $\alpha$ and $\lambda$ do not appear explicitly, this formula provides an expression
for $p$ in terms of the physical fields only,
\be p = \text{sign}(\Delta \chi_0) {q_{D1} Q_{D5} +Q_{D1} q_{D5} \over Q_{D1} q_{D5} - q_{D1} Q_{D5} } \sqrt{{ Q_{D5} (r_0^2 + r_\infty^2)^2  \over Q_{D1} \Delta \chi_0^2 }+1 }
\label{p-Phys} \ee
From the expression (\ref{p-Phys}), we see once more that we need a non-vanishing jump in the axion.
Since $p$ needs to stay finite, we need ${q_{D1} \over q_{D5}} \neq { Q_{D1} \over Q_{D5}}$.
This requirement is also expected from the analysis in Section \ref{secfiveb},
where we need different charge vectors $\hat n_k$ to avoid a vanishing angle between two strings.
Moreover, we know that $p$ needs to be positive to have a regular solution. This leads to the simple regularity requirement
\be { Q_{D1}^2 q^2_{D5} - q^2_{D1} Q^2_{D5} \over \Delta \chi_0} > 0  \label{Reg-Phys} \ee
which only involves the physical fields. We are left with two parameters to fix, $\alpha$ and $\lambda$.
It is convenient to write an expression for the ratio of the central charges
\be {q_{D1} q_{D5} \over Q_{D1} Q_{D5} } = {\alpha^2 \over 4} \Big( \lambda + {1 \over \lambda} \Big)
{\Big( p(r_0^2+r_1^2) + {\textstyle r_\infty^2- r_1^2  \over \textstyle p} \Big)^2 \over
(r_0^2 + r_\infty^2)^2 + {\textstyle Q_{D1} \over \textstyle Q_{D5}} \Delta \chi_0^2 }  \label{expr-alpha}   \ee
It is immediate to solve this expression  for $\alpha$. Finally, we can write the jump in the axion of the probe region, $\Delta \chi_1 = \chi_1 - \chi_\infty$, as follows,
\be \Delta \chi_1 = { { \textstyle q_{D5} } \over  p^2-1 } \left\{ {(r_0^2+r_1^2)p^2+ r_\infty^2 -r_1^2 \over 2 \sqrt{q_{D1} q_{D5}}}
\Big( \lambda - { 1 \over \lambda} \Big) +  {2 p^2 Q_{D1} \Delta \chi_0 \over q_{D1} Q_{D5} + Q_{D1} q_{D5} } \right\}   \ee
This is a simple quadratic equation and can be solved exactly,
\be \lambda = \Omega + \sqrt{\Omega^2+1}, \qquad  \Omega= \sqrt{q_{D1} \over q_{D5}} {  \displaystyle {   \textstyle (q_{D1} Q_{D5} + Q_{D1} q_{D5}) ( p^2 -1) \Delta \chi_1 }
-  {  \textstyle 2 p^2 Q_{D1} q_{D5}  } \Delta \chi_0  \over  \textstyle (q_{D1} Q_{D5} + Q_{D1} q_{D5}) \big( (r_0^2+r_1^2)p^2+ r_\infty^2 -r_1^2 \big)
 } \label{sol-lambda1} \ee
If we use Equation  (\ref{p-Phys}), the parameter $\lambda$ can be expressed only in terms of physical fields.
Similarly, the expression (\ref{sol-lambda1}) can be used with (\ref{expr-alpha}),  (\ref{expr-beta}), (\ref{expr-gamma}) and (\ref{expr-omega})
to express all the probe parameters in terms of the physical fields alone.

The expression for $\lambda$ (\ref{sol-lambda1}) is manifestly real and positive for all values of the physical parameters.
Moreover, according to (\ref{expr-alpha}), if $\lambda$ is positive, then $\alpha$ will be real and $A$ will have its singularities on the real axis.
It follows that the relation (\ref{Reg-Phys}) is the only constraint on the physical fields coming from regularity.

The final step is to consider the order $\epsilon$ corrections to the background global parameters.
After some algebra, one can see that the residues of $H$ get order $\epsilon$ corrections only through the constant $B_0$,
\be c_{H,0}= -p^2 {r_0^2 \over 2 B_0}, \qquad  c_{H,\infty}= - {r_\infty^2 \over 2 B_0}  \ee
The residues of $A$ become
\be
c_{A,1}+c_{A,2}= - { B_0 Q_{D5} \over 4 \pi^2} - p^2  \sqrt{\gamma} \Big( \lambda + {1 \over \lambda} \Big) \epsilon, \qquad
c_{A,1}-c_{A,2}=  p \Delta \chi_0  - p  \sqrt{\gamma} \Big( \lambda + {1 \over \lambda} \Big) \epsilon
\ee
The multiplicative constant $B_0$ has simple corrections as well
\be B_0 = - 4 \pi^2 p \sqrt{{(r_0^2+r_\infty^2)^2 \over Q_{D1} Q_{D5}} + {\Delta \chi_0^2 \over Q_{D5}^2} } \left( 1 + {\epsilon \over 2} \text{sign}\ (Q_{D1} q_{D5} - q_{D1} Q_{D5} ) \Big({q_{D1} \over Q_{D1}}+{q_{D5} \over Q_{D5}} \Big)  \right) \ee

With these expressions for the parameters the Neveu-Schwarz charges in all regions vanish up to order $\epsilon^2$.

\subsubsection{The doubly-degenerate limit}\label{Sec-doubledeg}

The probe limit studied in the previous subsection allowed us to compute the global parameters of our solution in terms of the physical
data in the asymptotic regions order by order in $\epsilon$. However, in that limit, the charges in one of the asymptotic regions
need to be small and the jump in the unattracted axion cannot be set to zero.
It is desirable to find a tractable class of solutions which have charges of order one in all asymptotic regions and display
jumps only in the six-dimensional dilaton. In this subsection, we examine another limit in which four poles of $A$ pairwise become very close to
two asymptotic region. We show that it leads to solutions with the above properties. However, there is a price to pay since
the expansion parameter $\epsilon$ can not be taken all the way to zero without getting a solution with some singularity.

If we want to keep the asymptotic regions at $z=0,1,\infty$, the harmonic function $H$ can be taken in the form
\be H = i \Big( \epsilon {c_{H,0} \over z} + \epsilon {c_{H,1} \over z-1} - z {c_{H,\infty } \over \epsilon} \Big) + c.c. \ee
and $A$ is equal to
\be A= i \Big( \epsilon {c_{A,1} \over z - \alpha \epsilon } + \epsilon {c_{A,2} \over z + \alpha \epsilon} + \epsilon {c_{A,3} \over z - 1- \beta \epsilon } + \epsilon {c_{A,4} \over z -1+ \beta \epsilon} + a \Big)  \ee
Here $\epsilon$ is taken to be small but finite.  The reasons why some of the residues have a factor of $\epsilon$ will be clear in the following
analysis. Also note that this is not the most general ansatz since the poles of $A$ approach the poles of $H$ at the same rate.
The function $B$ is taken to be equal to
\be B = i  {\Big( c_{H,0} (z-1)^2 + c_{H,1} z^2  \Big) \epsilon^2 + c_{H,\infty} z^2 (z-1)^2
\over (z^2 -\alpha^2 \epsilon^2) \big((z-1)^2 - \beta^2 \epsilon^2 \big) }  \ee
To obtain the above expression, the constant $B_0$ in the definition (\ref{bdefine}) of $B$  has been set to $-\epsilon$.
We can focus our analysis to the region of $\Sigma$  close to the pole of $H$ at $z=0$ and introduce the coordinates
 \be z \rightarrow \epsilon u  \ee
The functions $H$ and $A$ can be written in the new coordinates as
\bea H &=& i \Big( {c_{H,0} \over u} + {c_{H,1} \over u - { 1 \over \epsilon}} - c_{H,\infty} u \Big)  +c.c. \no \\
A &=&  i \Big( {c_{A,1} \over u-\alpha} +  {c_{A,2} \over u + \alpha } + {c_{A,3} \over u- {1\over \epsilon} - \beta } +
{c_{A,4} \over u- {1 \over \epsilon} + \beta } + a \Big)  \eea
This change of coordinates has moved one of the asymptotic regions from $z=1$ to  $u=1/\epsilon \gg 1$.

In the limit $\epsilon \rightarrow 0$,  $A$, $B$ and $H$ become equal to the functions corresponding to a strip solution
with only two asymptotic regions at $u=0$, $u= \infty$.
In analogy with the analysis of the strip solution carried out in \cite{Chiodaroli:2009yw}, it is possible to set the axion to zero
in the two asymptotic regions. Interestingly, the dependence from the asymptotic region at $z=1$ has completely dropped.
In other words, the regions at $z=0$ and $z=1$ lost communication and, as far as the region close to $z=0$ is concerned, there is only
another asymptotic region for $z=\infty$. Note that we cannot take the limit $u \rightarrow \infty $ in the above expressions if $\epsilon$
is small but finite: at some point, order $\epsilon$ terms would become relevant and we would start getting to the other
two asymptotic regions.

The analysis for the region close to $z=1$ is completely analogous. On the other side,
the portion of $\Sigma$ where $z$ is of order one can become singular, with the metric factor $f_1$ blowing up like $1 \over \epsilon$.
This makes intuitive sense since, if we get the degeneration limit all the way,  the two regions at $z=0,1$ are stretched infinitely far from the region at $z=\infty$.
Since all the poles of $A$ are located in those two regions, the pole of $H$ at $z = \infty$ is left alone and the solution is expected to become singular.
However, the singular region is stretched infinitely apart from the regions at $z=0,1$ in the degenerate limit.

In the following analysis, we will keep $\epsilon$ small but finite so that the solution is everywhere regular.
It is possible to express the global parameters in terms of the physical fields order by order in $\epsilon$.
We can focus on solutions with no jump in the unattracted axions and with vanishing Neveu-Schwarz charges.
With some algebra, one can express the residues of the harmonic function $H$ as
\bea c_{H,0} &=& { Q_{D1} Q_{D5} \over 32 \pi^4} {r_0^2 \over (r_0^2 + r_\infty^2)^2} + O(\epsilon^2 ) \no \\
c_{H,1} &=& { q_{D1} q_{D5} \over 32 \pi^4} {r_1^2 \over (r_1^2 + r_\infty^2)^2} + O(\epsilon^2 ) \no \\
 c_{H,\infty} &=& {  r_\infty^2  \over 2} + O(\epsilon^2 )   \eea
Note that there are no order $\epsilon$ corrections. The positions of the poles of $A$ are now given by
\be \alpha= {1 \over 4 \pi^2} {\sqrt{Q_{D1} Q_{D5}} \over r_0^2 + r_\infty^2 }+ O(\epsilon^2),
\qquad \beta= {1 \over 4 \pi^2} {\sqrt{q_{D1} q_{D5}} \over r_1^2 + r_\infty^2 } + O(\epsilon^2) \ee
The shift constants in the definition of $A$ and $K$ are of order $\epsilon$, while the expressions for the residues of $A$ to the leading order are,
\be c_{A,1}=c_{A,2}= {Q_{D5} \over 8 \pi^2}  + O(\epsilon),  \qquad
c_{A,3}= c_{A,4}={q_{D5} \over 8\pi^2} + O(\epsilon) \ee
To have regular solutions with positive residues, all charges need to have the same sign.
It is also interesting to note that the solution is singular if the ratios of the charges in the two regions are equal to each other, $q_{D1}/Q_{D1}=q_{D5}/Q_{D5} $.
Solutions of this sort are expected to be singular from the analysis in Section \ref{secfiveb}
since two branches of a string junction collapse when the corresponding unit norm vectors $\hat n_k$ become parallel.

\section{Boundary entropy}\label{holoent}
\setcounter{equation}{0}

A proposal to calculate the entanglement entropy of a $CFT_{d}$ with a dual description
as a gravitational theory in $AdS_{d+1}$ was discussed in \cite{Ryu:2006bv,Ryu:2006ef}. 

In Poincar\'{e} coordinates,
the CFT is defined on the Minkowski space $R^{1,d-1}$. This space can be
thought of as the boundary of $AdS_{d+1}$.
A subsystem ${\cal A}$ is a $d$-dimensional spatial region in a constant-time slice.
One can find a static minimal surface $\gamma_{\cal A}$ which extends into the $AdS_{d+1}$ bulk and ends on the boundary of  $ {\cal A}$
as one approaches the boundary of $AdS_{d+1}$.
The holographic entanglement entropy is then given by the following formula  \cite{Ryu:2006bv,Ryu:2006ef},
\bea
S_A = {{\rm Area}(\gamma_{\cal A}) \over 4 G^{(d+1)}_{N}}
\eea

where ${\rm Area}(\gamma_{\cal A})$ denotes the area of the minimal surface $\gamma_{\cal A}$ and $ G^{(d+1)}_{N}$ is
the Newton constant for $AdS_{d+1}$ gravity.
In the case of $AdS_3$, the area ${\cal A}$ is an interval and the boundary $\partial {\cal A}$ is a collection of points.
The minimal surface is a spacelike geodesic connecting these points in the $AdS_3$ bulk.

It was argued in   \cite{Calabrese:2004eu,Azeyanagi:2007qj}  that the boundary entropy of an interface conformal
field theory can be related to the entanglement entropy in case  ${\cal A}$ is a symmetric interval enclosing  the interface.
The holographic prescription was used in \cite{Azeyanagi:2007qj} to calculate boundary entropy for
the non-supersymmetric Janus solution with $AdS_3$ asymptotics,  finding agreement to leading order in the deformation parameter with
a boundary conformal field theory calculation.

In \cite{Chiodaroli:2010ur} the holographic calculation of the boundary entropy was generalized for the BPS Janus solution  with $AdS_3\times S^3$ asymptotics
 found in \cite{Chiodaroli:2009yw}.  The prescription for the minimal area surface has to be generalized. The minimal
 area surface is integrated over all of $\Sigma$ as well as $S^2\times M_4$.
 \bea
 S_A =  {1\over 4 G_N} \int_{S^2} d\Omega_2 \int_{M_4} d\Omega_4 \int_{\Sigma} \  \rho^2 f_2^2 f_3^4
\label{entanglement10d} \eea
\begin{figure}
\centering
\includegraphics[ scale=0.41]{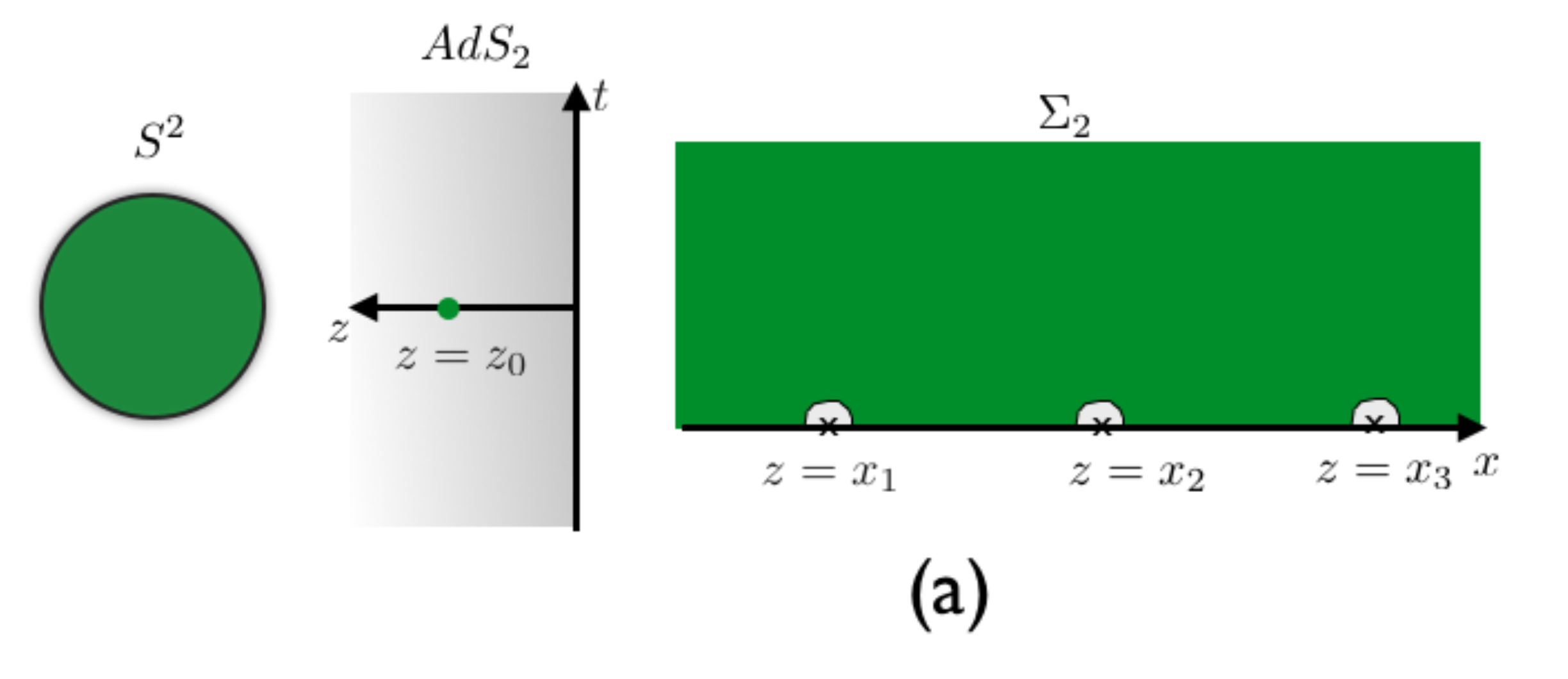}
\includegraphics[ scale=0.36]{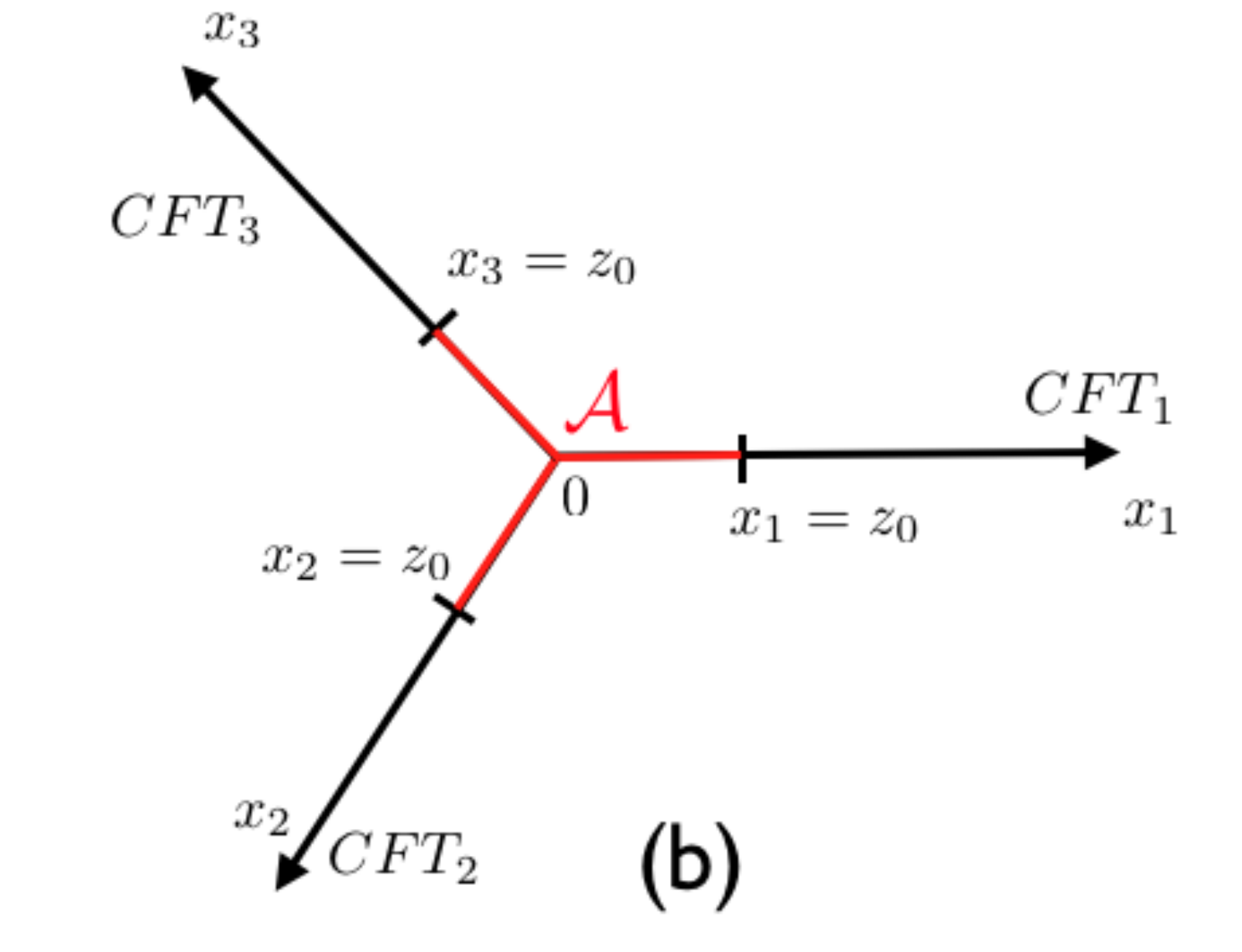}
\caption{{\bf (a)} The static minimal surface for entanglement entropy covers two-sphere $S^2$ and Riemann surface  $\Sigma$ and located at $z=z_0$ in $AdS_2$.  The UV cutoff is taken around the three poles of $H$. {\bf (b)} On the CFT side the area ${\cal A}$ is the star shaped region with interval $x_i \in [0,z_0]$ in the three leads $i=1,2,3$.}
\label{fig3}
\end{figure}
 Using this prescription it was shown in \cite{Chiodaroli:2010ur} that the holographic
 boundary entropy exactly agrees with the BCFT calculation in the case where the
 six-dimensional dilaton jumps across the interface. It must be noted that the integration over the fibered three-sphere
and the $M_4$ can be interpreted as  tracing over all KK-states.

\subsection{Holographic boundary entropy for three-junctions}
We now calculate the entanglement entropy for the star-shaped region depicted in Figure \ref{fig3}.
Our starting point is a relatively simple expression for the integrand in (\ref{entanglement10d}),
\be f^2_2 f^4_3 \rho^2 = {(A+\bar A) K - (B+\bar B)^2 \over |B|^2} |\partial H|^2
 = {4  \over B_0^2}  \left( \displaystyle \sum_{  i,j }  { L_{ij}  \over |z- p_i|^2 |z-p_j|^2 }
\right) { y^2 \displaystyle \prod^4_{i=1} |z-p_i|^2 \over |z|^4 |z-1|^4} \qquad \ee
where the matrix $L_{ij}$ is defined as
\be L_{ij} = {c_{A,i} c^2_{B,j} \over c_{A,j}}  -  c_{B,i} c_{B,j}  \ee
The entanglement entropy is then given by the integral of a rational function. It is convenient to expand this integral using a
basis of suitable integrals,
\be S_A = { V_{S^2} \over 4 G_N B^2_0} \big( l_1 I_1 + l_2 I_2  + l_3 I_3 + l_4 I_4 + l_5 I_5 + l_6 I_6 \big)  \ee
The integrals are
\bea  I_1 = \int { (x-1)(|z|^2-x ) y^2 dx dy  \over |z|^4 |z-1|^4 } && I_2 = - \int { x (|z|^2-x ) y^2 dx dy  \over |z|^4 |z-1|^4 } \no \\
 I_3 = \int {x(x-1) y^2 dx dy  \over |z|^4 |z-1|^4} &&  I_4 = - \int {(x -1) y^2 dx dy  \over |z|^4 |z-1|^4 } \no \\
  I_5 = \int { x  y^2 dx dy  \over |z|^4 |z-1|^4} &&  I_6 = \int {(|z|^2-x ) y^2 dx dy  \over |z|^2 |z-1|^4}  \eea
The six integrals above have been chosen so that they are not independent.
First of all, the first three integrals, $I_1, I_2 $ and $I_3$, can be related by a change of coordinates and have the same value which can be computed exactly,
\be I_1=I_2=I_3= - {\pi \over 8} \ee
Moreover, the last three integrals,  $I_4, I_5$ and $I_6$, are divergent, and each of them can be regularized introducing a single cutoff.
In particular, the divergent contribution to $I_6$ comes only from the pole at $z=\infty$ while $I_4$ and $I_5$ diverge
due to the poles at  $z=0$ and $z=1$ respectively. These integrals are related by a change of coordinates as well,
and are equal to each other as functions of the respective cutoffs,
\be I_4(\lambda)=I_5(\lambda)=I_6(\lambda) =  {\pi \over 2} \log {1 \over \lambda} + O(\lambda^2) \ee
The entanglement entropy depends on the parameters of the solution only through the expansion coefficients, which are given as follows,
\be \begin{array}{rcl} \displaystyle l_1 &=&  - 4 \displaystyle \sum_{ i \neq j \neq k \neq l } L_{ij} p^2_k \\
l_2 &=&  \displaystyle - 4 \sum_{i \neq j \neq k \neq l } L_{ij} (1- p_k)^2 + l_6  \\
l_3 &=&  \displaystyle 2  \sum_{i \neq j \neq k \neq l } L_{ij} (4p_k p_l -2 p_k-2p_l+1)  \end{array} \qquad
\begin{array}{rcl} \displaystyle l_4 &=& \displaystyle  2 \sum_{i \neq j \neq k \neq l } L_{ij} p^2_k p^2_l  \\
l_5 &=& \displaystyle 2 \sum_{i \neq j \neq k \neq l } L_{ij} (1-p_k)^2 (1-p_l)^2 \\
l_6 &=&  \displaystyle 2\sum_{i \neq j \neq k \neq l } L_{ij} \end{array} \\ \ee
It is useful to note that three of these coefficients are related to the $AdS_3$ radii in the three asymptotic regions given by
\be  l_4= { B_0^2 } R^4_{AdS_3,0} , \qquad   l_5 = { B_0^2  } R^4_{AdS_3,1}, \qquad  l_6= { B_0^2 } R^4_{AdS_3,\infty}  \label{coeff-Rads} \ee
In conclusion, we can express the entanglement entropy as
\be S_A = {V_{S^3} \over 4 G_N B_0^2}  \Big( l_4 \log {1 \over \lambda_1} + l_5
\log {1 \over \lambda_2 } + l_6 \log {1 \over \lambda_3} + {l_1 + l_2 + l_3  \over 4}  \Big)
\label{Amin} \ee
Following the analysis of \cite{Chiodaroli:2010ur}, we consider separately each asymptotic region, introduce a Fefferman-Graham coordinate system
\be ds^2 =R_{AdS_3}^2 {d \xi^2 - dt^2 + d \eta^2 \over \xi^2} + \dots \ee
and set the cutoff at $\xi=\epsilon$. Using the notation in Equation \ref{Amin}, the cutoff becomes 
\be  \lambda= |z-p_H|= {\epsilon \over z_0 } {  2 |b_0| \over \sqrt{a_1 k_1 - b^2_1}}  \ee
The entanglement entropy can then be written as \cite{Calabrese:2004eu}
\be S_A = {c_F \over 6} \log{ z_0 \over \epsilon} + S_{bdy}  \ee 
where $c_F$ is the central charge of the BCFT obtained by folding the conformal field theories on the three branches of the junction.
The boundary entropy is the finite part of the entanglement entropy, 
\bea S_{bdy} = {V_{S^3} \over 4 G_N} \left( \sum_{a=0,1,\infty} R^4_{AdS_3,a} \log {R^2_{AdS_3,a} \over 4   \ c_{H,a}}
+  {1 \over  B_0^2 } \sum_{i \neq j \neq k \neq l} L_{ij} \big(p_k- p_l \big)^2  \right) \label{Sbundy} \eea
As explained in \cite{Calabrese:2004eu}, the boundary entropy is identified up to a non-universal constant which does not depend 
on the presence of the boundary. The contribution to the boundary entropy from the divergent integrals $I_4, I_5$ and $I_6$ has a
simple expression in terms of the $AdS_3$ radii in the various regions.
These terms are analogous to the ones arising in the solution with two asymptotic regions discussed in \cite{Chiodaroli:2010ur}.
The last term in Equation (\ref{Sbundy}) is the result  of the finite integrals $I_1, I_2$ and $I_3$, and has a less immediate
physical interpretation.

Given the complexity to express the parameters in terms of physical ones, it is perhaps more illuminating to consider the degeneration limits as discussed in the previous section.
To study the boundary entropy in the probe limit, it is useful to introduce the quantities
\be N_B=  {Q_{D1} Q_{D5}  \over 4 G_N V_{S^3}}, \qquad \Delta N_B=  { (Q_{D1} q_{D5} +q_{D1} Q_{D5})  \over 4 G_N V_{S^3}} \epsilon  \ee
With this notation, the central charges of the regions at $z=0$ and $z=\infty$ are equal to ${2 \over 3} N_B$ and  ${2 \over 3} (N_B + \Delta N_B)$
respectively, while the numbers of bosons in the corresponding conformal field theories are $N_B$ and $N_B + \Delta N_B$.
In the case of the probe limit, the total boundary entropy can be written as
\be S_{\text{bdy}}= {N_B + \Delta N_B \over 4} \log{( r_0^2+r_\infty^2)^2 + {\textstyle Q_{D1} \over \textstyle Q_{D5}} \Delta \chi_0^2  \over 4 r_0^2 r_\infty^2}
- {\Delta N_B \over 4} \log { \sqrt {  Q_{D1} \over  Q_{D5}}  {\Delta \chi_0 \over 2 r_0^2 }} + \tilde c  \label{probe-Sbdy} \ee
$\tilde c$ is a constant which does not depend on the unattracted scalars,
\be
\tilde c={\Delta N_B\over4}  \Big(  \log {Q_{D1} q_{D5} + q_{D1} Q_{D5} \over Q_{D1} q_{D5} - q_{D1} Q_{D5}} + \text{sign}(Q_{D1} q_{D5} - q_{D1} Q_{D5}) +  {1+ \text{sign} \ q_{D5} \over 2} \Big)  \label{ctildeone}
\ee
The order one terms in the above expressions give exactly the result for the two-pole solutions studied in \cite{Chiodaroli:2010ur}.
It is interesting to note that the scalars related to the probe asymptotic region, $r_1^2$ and $\Delta \chi_1$,
do not appear in any order $\epsilon$ correction.
This occurs because the radius of the probe region is of order $\epsilon^2$.
Instead, the leading correction to the boundary entropy comes from the change of central charge between
the two asymptotic regions at $z=0,\infty$
which is produced by the probe.
Moreover, the combination $\sqrt{Q_{D1} \over Q_{D5}} \Delta \chi_0$ is exactly the unattracted axion in the region at $z=0$.
Hence, the above expression depends on the $D1$- and $D5$-charges only through $N_B$ and $\Delta N_B$ while the ratios of charges
contribute only to an additive constant that, in principle, can be set to zero with a redefinition of the cutoff.
As expected, the boundary entropy depends on the unattracted scalars logarithmically.
The order $\epsilon$ correction involving the unattracted axion has the same form as the one involving $r_0$.

\medskip

On the other hand, the boundary entropy in the doubly-degenerate limit evaluates to
\be
 S_{\text{bdy}}={N_{B,0} \over 4} \log{(r_\infty^2+r_0^2 )^2 \over  4 r_0^2 r_\infty^2}+
{N_{B,1} \over 4} \log{(r_\infty^2+r_1^2 )^2 \over4 r_1^2 r_\infty^2} +{N_{B,1}+N_{B,0}-N_{B,\infty} \over 4} \log{2 r_\infty^2}
+ \tilde c  \label{probe-dubdeg}
   \ee
where we denote here the number of compact bosons in each asymptotic region respectively by $N_{B,0}$, $N_{B,1}$ and $N_{B,\infty}$.
Again $\tilde c$ depends only on the charges and is equal to
\be
\tilde c = {N_{B,\infty} \over 8} \Big( \log{ \scriptstyle (q_{D1}+Q_{D1})(q_{D5}+Q_{D5}) \over 16 \pi^4} + 1 \Big) - { N_{B,0} \over 8} \Big( \log{Q_{D1} Q_{D5}\over 16 \pi^4} + 1 \Big) -{ N_{B,1} \over 8} \Big( \log{q_{D1} q_{D5} \over 16 \pi^4} + 1 \Big)
\label{ctildetwo}
    \ee

 The logarithmic term in the boundary entropy (\ref{probe-dubdeg}) allows for a simple interpretation  as the first two terms are the the same as the boundary entropy of two decoupled Janus interfaces with $N_{B,0}$ and $N_{B,1}$ bosons respectively.  The third term in(\ref{probe-dubdeg}) can be interpreted as coming from $N_{B,\infty}-N_{B,0}-N_{B,1}$ bosons which satisfy Neumann boundary conditions.

 Lastly, the non-logarithmic terms $\tilde c$ in both degeneration limits (\ref{ctildeone}) and (\ref{ctildetwo}) only depend on the charges but not the unattracted scalars. If one considers the difference of the boundary entropy of two configuration with the same
 charges but different jumps in the unattracted scalars, then $\tilde c$ will drop out of the
 expression. This may suggest that the term could be identified with the non-universal constant in \cite{Calabrese:2004eu}.
 We have however been unable to prove this statement away from the degeneration limit.

\subsection{Perspectives from a CFT toy model}\label{CFTbound}

The CFT theory dual to the $AdS_3\times S^3$ type IIB Supergravity solutions is a $\mathcal{N}=(4,4)$ CFT.
For generic values of the D1-,F1-,D5- and NS5-charges, this CFT is a U-dual of the D1/D5 CFT.
In this section we will focus on a much simpler toy model CFT, where the conformal boundary conditions for junctions and the boundary entropy can be calculated. The aim is to compare general features of  the CFT and holographic boundary entropy.

Under the \emph{folding trick}, one could fold the branches to the same side of the interface,
and treat the theory as a boundary CFT whose central charge is the sum of the central charges of all the branches.  Different interface CFT's then correspond to different conformal boundary conditions in the folded theory.
The toy model theory we consider is  the action of $n$ free scalar fields defined on the half-line which is given by
\be
S= \int _{\sigma>0}d\tau d\sigma \sum_{i,j} g_{ij} \Big( \partial_\tau \phi_i \partial_\tau \phi_j -  \partial_\sigma \phi_i \partial_\sigma \phi_j\Big)
\ee
and we take $g_{ij} = \delta_{ij}$.
In the context of quantum wires which are usually systems of fermions, the bosonic description of interest here is obtained via bosonization.
The currents $\partial_{\pm} \phi_i$, where $\sigma_\pm = \sigma\pm\tau$, are related to the chiral fermion numbers which are usually
taken as discrete \cite{Affleck:1991tk}. It is therefore most natural for us to consider compact scalars, each having radius $R_i$
\be
\phi_i \sim \phi_i + 2\pi R_i, \quad \quad i=1,2,\cdots n
\ee
Note that the metric we have taken is not the most general. If one imposes that all the scalars have the same periodicity, one rescales
the scalars so that the metric would become $g_{ij}={\rm diag}(R_1^2,\cdots R_n^2)$.
The stress energy tensor of the theory is given by
\be
T_{\sigma\tau} =\sum_{i=1}^n \partial_\tau \phi_i \partial_\sigma \phi_i
\ee
The variation of the action picks up a boundary term
\be\label{boundt}
\delta S_{\rm bound} = \int d\tau \sum_{i=1}^n \partial_\sigma \phi_i  \delta \phi_i |_{\sigma=0}
\ee
Energy momentum conservation enforces that
\bea\label{emtensor}
T_{\sigma\tau}|_{\sigma=0} =  \sum_{i=1}^n \partial_\tau \phi_i \partial_\sigma \phi_i|_{\sigma=0}=0
\eea
The variational principle demands that the boundary term in (\ref{boundt}) vanishes at $\sigma=0$.

A simple solution of this problem preserving time reversal and parity is given by imposing Neumann or Dirichlet boundary conditions
on the $\phi_i$. More generally we can have
\be
\hat \phi_i(\sigma,\tau)  = O_j^{\; i} \phi_j (\sigma, \tau)
\ee
with $O\in SO(n)$. We impose
\bea\label{boundcona}
\partial_\sigma \hat \phi_i |_{\sigma=0}& =&0 , \quad \quad {i=1,2,\cdots p} \no\\
\partial_\tau \hat \phi_i |_{\sigma=0}& =&0 , \quad \quad {i=p+1,\cdots n}
\eea
Since the conditions (\ref{boundt}) and (\ref{emtensor}) are invariant under $SO(n)$ rotations they are indeed solved by the above conditions. In terms of the original fields, (\ref{boundcona}) translates to
\bea\label{boundconb}
\sum_{j=1}^n O_{i}^{\; j} \partial _\sigma \phi_j &=& 0, \quad   {i=1,2,\cdots p} \no\\
\sum_{j=1}^n O_{i}^{\; j} \partial _\tau \phi_j &=& 0, \quad   {i=p+1,\cdots n}
\eea
It is also convenient to define the matrix
\be
S = O^T \left(\begin{array}{cc}  \mathbb{I}_p & \\ & -\mathbb{I}_{n-p} \end{array}\right) O
\ee
such that the boundary conditions can be re-written as
\be
\partial_+ \hat{\phi}_i |_{\sigma=0} = \sum_{j=1}^n S_i^j \partial_- \hat{\phi}_j |_{\sigma=0}
\ee
as in \cite{Bellazzini:2006kh, Bellazzini:2008cs}, which have discussed them in detail.
We notice that these boundary conditions are well-known in the context of open-string theory. 
They describe precisely D$p$-branes wrapping $p$-cycles
in an $n$-torus.
It is interesting to note that in \cite{Bellazzini:2006kh, Bellazzini:2008cs}, the scalars considered are generally non-compact. However, for compact scalars of given radii, these boundary conditions only preserve non-trivial zero-modes at discrete sets of $\{O_{i}^{\;j}\}$, corresponding to
D$p$-branes wrapping the torus only a finite number of times.

\begin{figure}
\centering
\includegraphics[scale=0.40]{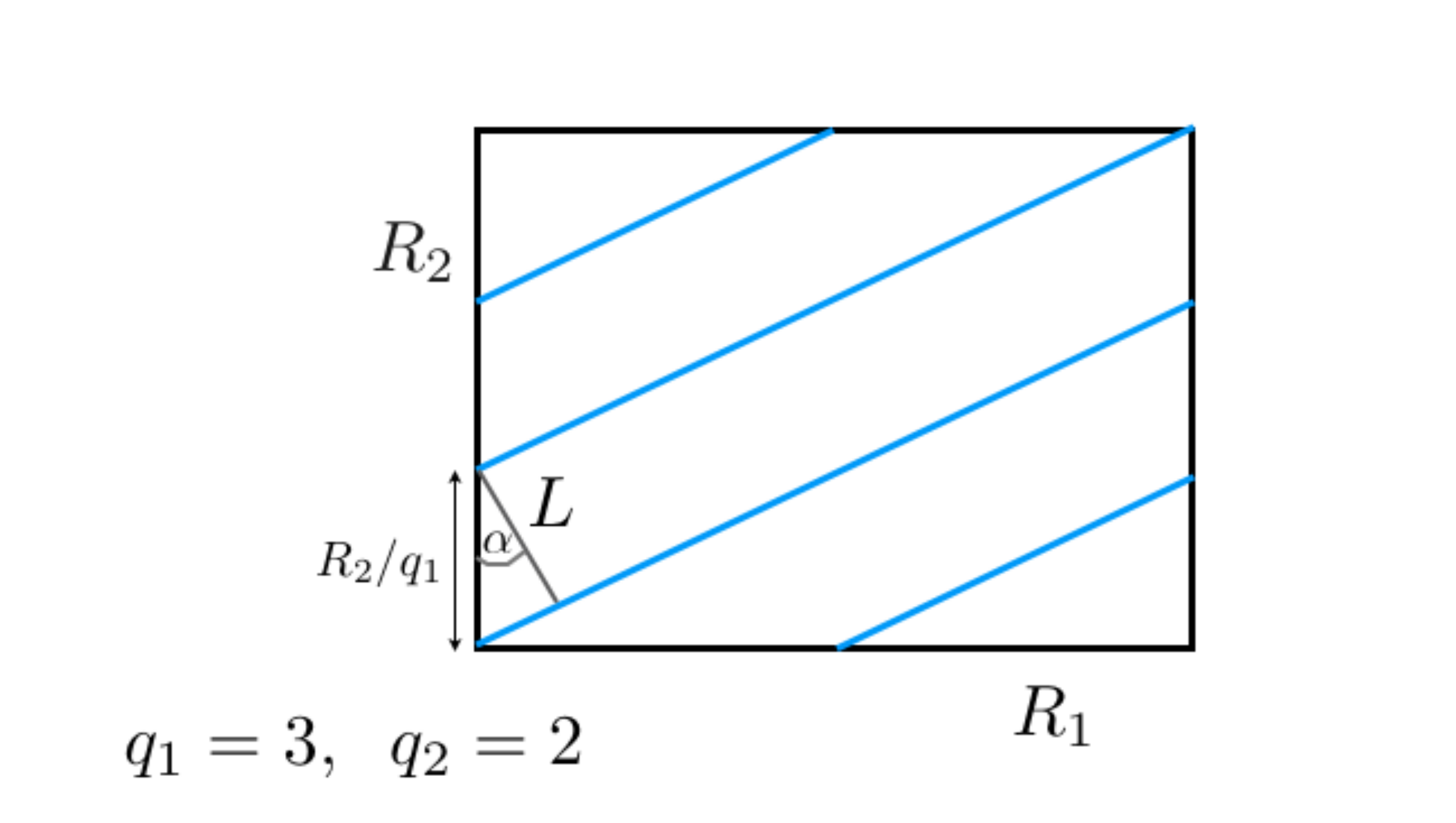}
\caption{wrapping of a $D1$-brane with $q_1=3, q_2=2$.}
\label{fig:3}
\end{figure}

In the case of $n=2$, i.e. 2 branches joined at the interface, each with a free boson, the most general boundary conditions preserving non-trivial zero-modes correspond to a $D1$-brane
with winding numbers $q_1$ and $q_2$, which are relatively prime integers, around the respective sides of the torus. An example is shown in Figure (\ref{fig:3}).

The boundary condition is given by
\be
\mathcal{O} =\left( \begin{array}{cc} \cos \alpha & \sin \alpha \\
-\sin \alpha & \cos \alpha
\end{array}\right)
\ee
with
\be\label{choiceal}
\cos\alpha= { q_1 R_1\over \sqrt{ (q_1 R_1)^2+ (q_2R_2)^2}}, \quad \sin\alpha= { q_2 R_2\over \sqrt{ (q_1 R_1)^2+ (q_2 R_2)^2}}
\ee
The corresponding boundary entropy is $S_{\textrm{bdy}}=\log g_B$, where
\be
g_B= \sqrt{  q_1^2 R_1^2 +q_2^2 R_2^2\over 2 R_1 R_2}
\ee
The holographic calculation of the boundary entropy matches exactly the above free field BCFT calculation \cite{Chiodaroli:2010ur}, with winding $q_1=q_2=1$.
It is remarkable that the supergravity solution picks out the precise boundary condition corresponding to a $D1$-brane wrapping the diagonal of the torus exactly once.

In the case of $n=3$, non-trivial boundary conditions could be obtained by considering either $D1$- or $D2$-branes. It is not clear if there are simple
expressions for the most general windings that allow non-trivial zero-modes. However, guided by the geometric picture it is not difficult to find explicit examples.
The most immediate generalization of the $n=2$ case would be for a $D1$-brane to wrap $q_1$ and $q_2$ times around two of the three-cycles, with the third cycle satisfying simple Dirichlet boundary condition. The $g_B$ factor is then given by
\be
g_B= \sqrt{  q_1^2 R_1^2 +q_2^2 R_2^2\over 2 R_1 R_2 R_3}
\ee
One could also consider a $D1$-brane wrapping the diagonal of the 3-torus. i.e. wrapping once around each cycle. The boundary condition is given by
the $S$-matrix
\be\label{toythree}
S = \frac{1}{R_1^2+R_2^2+R_3^2}\left(\begin{array}{ccc}R_1^2-R_2^2-R_3^2 & 2 R_1 R_2 & 2 R_1 R_3 \\
2 R_1 R_2  & -R_1^2+R_2^2 -R_3^2& 2 R_2 R_3 \\
2 R_1 R_3 & 2 R_2 R_3 & -R_1^2 - R_2^2 + R_3^2\end{array}\right)
\ee
and the corresponding $g_B$ factor is given by
\be
g_B= \sqrt{R_1^2 +R_2^2+ R_3^2 \over 2 R_1 R_2 R_3}
\ee

In all of these examples with non-trivial boundary entropy corresponding to the brane wrapping a finite number of times in the $n$-torus, the value of the boundary entropy is always given by
\be
g = \frac{V_{p}}{\sqrt{2^{p} V_{T^{n}}}}
\ee
where $V_{p}$ is the volume of the D$p$-brane and $V_{T^{n}}$ is the volume of the $n$-torus.

Comparing the result of the toy model three CFT junction  (\ref{toythree}) with the supergravity calculation (\ref{Sbundy}), it is interesting to note that for the supergravity result  there is a constant term appearing in addition to logarithm.

However, the log terms in the supergravity result does share many features with the toy model. To begin with, let us revisit the probe limit.
We can read off the $g_B$ factor there
\be g_B= { \Big( ( r_0^2+r_\infty^2)^2 + {Q_{D1} \over Q_{D5}} \Delta \chi_0^2 \Big)^{N_B + \Delta N_B \over 4} \over \sqrt{2^{N_B}r^{N_B }_0 r^{N_B + \Delta N_B }_\infty \big( 2\sqrt{  Q_{D1} \over  Q_{D5}}  \Delta \chi_0 \big)^{\Delta N_B\over 2}} }
 \label{probe-g} \ee
If we consider only the dependence of (\ref{probe-g}) from the six-dimensional dilaton, we see that
the numerator of the $g_B$ factor can be interpreted again as the volume of a $N_B+\Delta N_B$-dimensional brane,
while the denominator gives the square root of the volume of a $2N_B+{ \Delta N_B}$ dimensional torus\footnote{There is a small mismatch in factors of 2 but that is only of order $\Delta N_B$.}. In fact, to zeroth order in $\Delta N_B$, the contribution including the axion matches with the CFT toy model in the presence of world-volume $B$-fields, as already noted in \cite{Chiodaroli:2010ur}, where the axion plays the role of $B$-fields.

We have also noted that for each given set of physical charges and moduli, the supergravity solution can be realized by more
than a set of parameters, due to the non-linear relation between them.
It suggests that the supergravity solution might be realizing different boundary conditions across the jumps, i.e. different winding numbers.

Note however that due to the complexity of the CFTs  at this point it seems impossible to construct a precise map between the supergravity solutions and the exact conformal boundary conditions on the junction. The interpretation of the non-logarithmic terms in (\ref{Sbundy}) is not clear. It is interesting to note
however, that the term, at least in the degenerate limits, depends purely on the central charges. It is reminiscent
of the non-universal terms as briefly discussed in \cite{Calabrese:2004eu}, which does not vary with boundary conditions for a given theory.

\section{Discussion}\label{discus}
\setcounter{equation}{0}

In this paper  we have continued the analysis of the half-BPS solutions  found in \cite{Chiodaroli:2009yw} which are locally asymptotic to $AdS_3\times S^3\times M_4$. In particular,  we have derived formulae for all Page charges associated with the solution, discussed the relevance of
supersymmetry and the attractor mechanism and clarified the relation of the moduli of the
solution and the physical parameters, i.e.  the Page charges and asymptotic values of unattracted scalars.

The discussion of the attractor mechanism also leads to a conjectured  relation between the half-BPS supergravity solution and
 a junction of self-dual BPS strings in six flat dimensions.
A single half-BPS dyonic string solution in six-dimensional supergravity with
D1, F1, NS5 and D5 charges  produces an $AdS_3 \times S^3$ supersymmetric vacuum the near-horizon limit. The decoupling
limit of the worldvolume theory produces a $\mathcal{N}=(4,4)$
two-dimensional superconformal field theory and the two theories are dual via the AdS/CFT
correspondence.  One  would expect that away from the string junction the decoupling limit is
still at work. As argued in Section \ref{secfiveb} the string junction in flat space preserves four
of the sixteen supersymmetries of $\mathcal{N}=(2,0)$ supergravity. The decoupling limit is expected to
enhance the supersymmetries from four to eight, which is the number of preserved
supersymmetries of the  half-BPS interface solutions.
Moreover, the solutions we have found become singular when the charge vectors in two asymptotic regions are taken to be parallel.
Correspondingly, in case of parallel charges, the angle between two strings in a six-dimensional junction goes to zero and
the two branches collapse.
In this paper, we have discussed the half-BPS interface solution with three asymptotic regions in detail.
It is straightforward to generalize the construction to solutions with $n$ asymptotic regions and to relate them to
junctions of $n$ dyonic strings
in six dimensions.

We have explored the holographic dual of the supergravity solutions which is given by a junction of $n$ CFTs.
The CFT is determined both by the four brane charges  and the  two unattracted scalars of  a given asymptotic region. The charges determine which particular U-
duality transformation of the well known $D1/D5$  CFT  one considers in each boundary
component. The values of the unattracted scalars determine at which point of the moduli space, with respect to the two
marginal deformations, the CFT lies.
Additional evidence for the identification between supergravity solutions and  interface CFTs is provided by the exact agreement of
some calculations in the two theories.
 In \cite{Chiodaroli:2010ur}  it was shown that the holographic interface entropy of the half-BPS Janus solution
constructed in \cite{Chiodaroli:2009yw} exactly agrees with the boundary  entropy  of the corresponding interface CFT.
In the present paper we have generalized the calculation of the holographic boundary entropy to the case of a junction of three CFTs.
While the result (\ref{Sbundy}) shares some features of the boundary entropy of a toy model CFT introduced in  Section \ref{CFTbound},
there are important differences. In particular the appearance of a non-logarithmic term in the holographic boundary entropy has no
counterpart in the toy model calculation. The CFTs of the BPS junction are however much more complicated than the single
compact bosons considered in the toy model. First, for generic values  of the parameters the CFTs are not D1/D5 CFTs, which are relatively
well understood. Second, since the central charges are expressed in terms of conserved charges, the toy model for the junction where all
three central charges are equal seems impossible to realize. It would be very interesting to find an interpretation for these terms on the CFT side. One suggestion is that they are related to the tripartite entanglement (see e.g. \cite{Coffman:1999jd}) of the CFT junctions. One should note however, that the degenerate limits suggest that these non-log terms depend only on the charges, and not on the unattracted scalars, which, in this respect, bears some resemblance to
 non-universal terms in the boundary entropy as in \cite{Calabrese:2004eu}.

Due to the  highly nonlinear nature of the expressions for the charges and non attracted  scalars, it is most likely impossible to invert
these relations in closed form, i.e. express the moduli of the solution in terms of the charges and the values of non attracted scalars in
the asymptotic region. In order to simplify the discussion we considered several degeneration limits where poles of the meromorphic
functions approach each other and residues scale in a prescribed fashion\footnote{
See also  \cite{Chiodaroli:2009xh} for a similar discussion involving higher genus Riemann surfaces.}.

These limits have provided us with tractable classes of solutions where the global moduli have simple expressions in terms of the physical parameters.
It would be very interesting to use these solutions to calculate quantities - such as reflection and transmission coefficients and bulk-boundary
correlation functions - in the dual interface theories.

The junctions we have constructed have the form of a so-called star-graph, i.e. $n$ semi infinite lines joined at a single point.
For 1+1-dimensional CFTs, more general configurations are possible.
For example instead of a triple junction one could consider two neighboring interfaces
on the real line, where the  CFT in the middle only lives in a finite interval.
Such configurations are important for considering the fusion of interfaces on the CFT side (see e.g  \cite{Bachas:2007td,Bachas:2008jd}).
Further investigation is required to determine whether such configurations can be realized holographically, possibly  by considering more general metrics than $AdS_2$.

A promising possible application of the duality between our supergravity solutions and CFTs on star graphs is
given by the study of quantum wires.
Electrons in one-dimensional conductors are not described by a Fermi liquid but instead by a Tomonaga-Luttinger (TL) liquid.
The spinless TL liquid has a bosonized formulation which corresponds to free compactified bosons.
The junction of two quantum wires has been analyzed utilizing conformal field theory techniques
(see for example \cite{Wong:1994pa}).
While the field theories associated with dyonic strings are more complicated than the TL liquids, which are also non supersymmetric,
one might hope that the qualitative features of the theories are similar. For example the discussion of multiple $(p,q)$ string
junctions given in \cite{Callan:1998sf} is very similar to the analysis
of the junctions of quantum wires in \cite{Bellazzini:2006kh,Bellazzini:2008cs,Oshikawa:2005fh}.
It is reasonable to expect that some of the properties of the strongly-coupled wires are modeled by the  dual holographic interface solution
in supergravity, which has been the case in other applications of the AdS/CFT correspondence
to the quark-gluon plasma and to higher-dimensional condensed matter systems.

\newpage

\bigskip

\noindent{\Large \bf Acknowledgements}

\medskip

We are grateful to C.~Bachas, I.~Brunner, E.~D'Hoker, J.~Estes, P.~Kraus, S.~Mathur and B.~Shieh for useful conversations.
The work of MG and MC was
supported in part by NSF grant PHY-07-57702.
The work of MC was supported in part by the 2009-10 Siegfried W. Ulmer Dissertation Year Fellowship and in part by NSF grant PHY-08-55356.
The research  of LYH at  the Perimeter Institute is supported by the Government of Canada
through Industry Canada and by the Province of Ontario through the Ministry of Research
and Innovation.
The research of DK is supported in part by the FWO - Vlaanderen, project G.0235.05
and in part by the Federal Office for Scientific, Technical and Cultural Affairs
through the ‘Interuniversity Attraction Poles Programme – Belgian Science
Policy’ P6/11-P.

\appendix

\section{Details on the charges}\label{appendixa}

In this appendix  we define the three kind of charges which can be associated with branes, namely the brane source charge, the Maxwell charge and the Page charge. For a detailed discussion of these charges see \cite{Marolf:2000cb}.
 The Bianchi identities for the AST are given by \footnote{Note that we use "supergravity" normalization of the $C_4$ instead of "string" normalization, which differ by a factor of 4.}
  \bea
  d\tilde F_3+d\chi\wedge H_3 \nonumber& =& 0 \\
  dH_3&=&0
  \eea
  Where
  \be\label{fthreedef}
  \tilde F_ 3 = dC_2 -\chi H_3, \quad \tilde F_5= dC_4- {1\over 4} C_2\wedge H_3
  \ee
 The equations of motion for the AST fields are
  \bea\label{eqofast3}
  d (e^\phi *\tilde F_3)-4  F_5\wedge H_3 &=& 0\nonumber\\
  d(e^{-\phi }* H_3)- e^\phi d\chi\wedge * \tilde F_3+ 4 F_5 \wedge \tilde F_3&=&0
  \eea
 In the following analysis we will only be interested in the five-brane and one-brane charge
 and do not display the expressions for the three brane charge.
  \subsection{Brane source charges}
    The brane source charge is defined by writing the Bianchi identity and equations of motion for the AST with all the terms on the left hand side. In the absence of brane sources the right hand side is zero, whereas if brane sources are present there are delta function sources on the right hand side. These sources  arise from the variation of the worldvolume action of the branes with respect to the antisymmetric tensor potential. Hence integrating the the left hand side defines the brane source charge.  The brane charge is localized and gauge invariant but neither quantized nor conserved.
     The brane source charges for the five branes and one branes are given by
      \bea
Q_{D5}^{bs} &=&\int_{V_4} \Big(d \tilde F_3 +d\chi\wedge H_3\Big)\no\\
Q_{NS5}^{bs} &=&\int_{V_4} dH_3 \no\\
Q_{D1}^{bs}&=&- \int_{V_8} \Big( d*\tilde F_3+ H_3\wedge  F_5\Big)\no\\
 Q_{F1}^{bs}&=& -\int_{V_8} \Big( de^{-\phi}*\tilde H_3-e^\phi d\chi \wedge *\tilde F_3 + 4   F_5\wedge \tilde F_3\Big)
  \eea
 \subsection{Maxwell charges:}
The Maxwell charges are given by the Integrals of the gauge invariant field strengths  for the fivebrane charge or their duals for the one brane charge
  \bea\label{maxcharge}
Q_{D5}^{max} &=&\int_{S^3} \tilde F_3, \qquad\quad\;\;
Q_{NS5}^{max} =  \int_{S^3} H_3, \no\\
Q_{D1}^{max} &=&-\int_{S^7} * \tilde F_3,  \qquad
Q_{F1}^{max} =-  \int_{S^7} e^{-\phi}*H_3
\eea
The Maxwell charges are gauge invariant and conserved but neither localized nor quantized.
   \subsection{Page charges}
  The page charges  can be obtained by writing the Bianchi identities as well as the equations of motion as $d j =0$ and integrating $j$ over the appropriate surface.  The Page charges are localized, quantized and conserved but are not gauge invariant.
   The Page charges for the five branes and one branes are given by
  \bea
  Q_{D5}^{Page} &=&\int_{S^3} \Big(\tilde F_3 +\chi H_3\Big)\no\\
Q_{NS5}^{Page} &=&  \int_{S^3} H_3 \no\\
Q_{D1}^{Page} &=&-\int_{S^7}\Big( e^\phi *  \tilde F_3 - 4C_4\wedge H_3\Big) \no\\
Q_{F1}^{Page} &=& - \int_{S^7} \Big( e^{-\phi} * H_3 -\chi e^\phi *\tilde F_3 +4 C_4\wedge dC_2 \Big) \eea
The formula for $Q_{F1}^{Page}$ follows from the fact that
the equation of motion  for $H_3$ (\ref{eqofast3}) can be rewritten in the following way:
\be
d\Big( e^{-\phi} * H_3 -\chi e^\phi *\tilde F +4 C_4\wedge dC_2\Big)=0
\ee

\subsection{Evaluation of the page charges}\label{pageeval}

The  ansatz for the complex rank three anti-symmetric  tensor field is  given by
\be
G= g^{(1)}_{a} f_1^2 \;e^{a}\wedge \omega_{AdS_2}+ g^{(2)}_{a} f_2 ^2\; e^{a}\wedge \omega_{S^2}
\ee
It  can be expressed in terms of potentials
\bea f_1^2 \rho e^{\phi/2} \Re(g^{(1)})_z & = & \partial_w b^{(1)} \label{potdef1}\\
 f_2^2 \rho e^{\phi/2} \Re(g^{(2)})_z & = & \partial_w b^{(2)} \label{potdef2b}\\
f_1^2 \rho e^{-\phi/2} \Im(g^{(1)})_z + \chi f_1^2 \rho e^{\phi/2} \Re(g^{(1)})_z  & = & \partial_w c^{(1)} \label{potdef3}\\
f_2^2 \rho e^{-\phi/2} \Im(g^{(2)}) _z+ \chi f_2^2 \rho e^{\phi/2} \Re(g^{(2)})_z & = & \partial_w c^{(2)} \label{potdef4}\eea
The potentials $c^{(1,2)}$ and $b^{(1,2)}$ are  expressed in terms of the meromorphic and harmonic functions as follows
\bea b^{(1)} &=& - {H (B + \bar B) \over (A + \bar A) K - (B + \bar B)^2 } - h_1, \qquad h_1={1 \over 2} \int {\partial_w H \over B} + c.c. \label{potharmonic1app}\\
 b^{(2)} &=& -i  {H (B - \bar B) \over (A + \bar A) K - (B - \bar B)^2 } + \tilde h_1, \qquad  \tilde h_1={1 \over 2 i} \int {\partial_w H \over B} + c.c. \label{potharmonic2app}\\
c^{(1)} & = & - i {H (A \bar B -  \bar A B) \over (A + \bar A) K - (B + \bar B)^2 } + \tilde h_2, \qquad \tilde h_2={1 \over 2 i} \int {A \over B}\partial_w H + c.c.  \label{potharmonic3app}\\
c^{(2)} & = & - {H (A \bar B +  \bar A B) \over (A + \bar A) K - (B - \bar B)^2 } + h_2, \qquad  h_2={1 \over 2 } \int {A \over B}\partial_w H + c.c.  \label{potharmonic4app}\eea
Using the expressions for the AST fields and the metric fields the expressions for the charges can be reduced to line integrals over a curve ${\cal C}$ which together with the $S^2$ forms a homology three-sphere.

\subsubsection{Five brane charges}\label{page5ch}
 In terms of the fields  in the paper we have
\bea
H_3= e^{\phi/2} Re(G), \quad \tilde F_3= e^{-\phi/2} Im(G)
\eea
and for the relevant charges one has for the NS5-brane
\bea
Q_{NS5}^{Page}&=& \int_{S^3} H_3
= \int_{S^3} e^{\phi/2} Re(G) \no\\
&=&  \int_{S^2} f_2^2 \rho e^{\phi/2} \big( Re(g^{2})_z dz+ Re(g^{2})_{\bar z} d\bar z\big) \no\\
&=& Vol(S^2) \Big(\int dz \; \partial_z b^{(2)} +\int d\bar z  \;\partial_{\bar z} b^{(2)} \Big)
\eea
and the Page charge for the D5-brane is
\bea
Q_{D5}^{Page} &=&\int_{S^3} \Big(\tilde F_3 +\chi H_3\Big)
=  \int_{S^3}  \Big(e^{-\phi/2} Im(G)   +\chi  e^{\phi/2} Re(G) \Big) \no\\
&=&  \int_{S^3} f_2^2  \rho \Big(e^{-\phi/2} \big(Im(g^{2})_z dz+ Im(g^{2})_{\bar z} d\bar z\big) +\chi e^{\phi/2} \big( Re(g^{2})_z dz+ Re(g^{2})_{\bar z} d\bar z\big) \Big)\no\\
&=& Vol(S^2)\Big(\int dz \; \partial_z c^{(2)} +\int d\bar z  \;\partial_{\bar z} c^{(2)} \Big)
\eea

 \subsubsection{ D1-brane charge}\label{paged1ch}
 In this section we evaluate the D1 brane charge as well as the fundamental string charge.
 For the D1-brane Page charge one has
 \bea\label{paged1}
 Q_{D1}^{Page,(b)} &=&-\int_{S^7}\Big( e^{\phi}* \tilde F_3 -4 C_4\wedge H_3\Big)
 \eea

 Where the first term in (\ref{paged1}) can be expressed as
  \bea
 \int_{S^7}  e^\phi * \tilde F_3 &=& \int_{M_4\times S^3}   e^{\phi/2} * Im(G) \no\\
 &=& \int_{M_4\times S^3}   e^{\phi/2} \Big(- i \;Im(g^{(1)}) _{z} e^{z234567}  + i \;Im(g^{(1)}) _{\bar z} e^{\bar z234567}  \Big)\no\\
  &=& \int_{M_4\times S^3}   e^{\phi/2} f_3^4 f_2^2 \rho \Big( -i \;Im(g^{(1)}) _{z} dz +i \;Im(g^{(1)}) _{\bar z}d\bar{z}\Big) \hat{e}^{234567}\no\\
   &=& Vol(S^2) \int  e^{\phi/2} f_3^4 f_2^2 \rho  \Big( -i \;Im(g^{(1)}) _{z} dz + i \;Im(g^{(1)}) _{\bar z}d\bar{z}\Big)\no \\
 &=& Vol(S^2) \int  e^{\phi} {f_3^4 f_2^2  \over f_1^2}  \Big( -i \big( \partial_z c^{(1)} -\chi \partial_z b^{(1)} \big) dz + i \big( \partial_{\bar z} c^{(1)} -\chi \partial_{\bar z} b^{(1)} \big) d\bar z \Big) \no\\
   &=& Vol(S^2)  \int   {4 K \over A+\bar A}  {(A+\bar A)K-(B+\bar B)^2\over (A+\bar A)K -(B-\bar B)^2}\Big( -i \big( \partial_z c^{(1)} -\chi \partial_z b^{(1)} \big) dz\no\\
   && + i \big( \partial_{\bar z} c^{(1)} -\chi \partial_{\bar z} b^{(1)} \big) d\bar z \Big) \no\\
    \eea
In the fifth line we used
  \bea
Im(g^{(1)}) _{z} = {e^{\phi/2}\over f_1^2 \rho} \partial_w c^{(1)} - {e^{\phi/2}\over f_1^2 \rho} \chi \partial_w b^{(1)}
\eea
The second term in (\ref{paged1}) can be expressed as
\bea\label{cfoura}
 \int_{M_4\times S^3}  4 C_4\wedge H_3&=&  \int_{M_4\times S^3}  4 C_4 \wedge e^{\phi/2} Re(G) \no\\
 &=&   \int_{M_4\times S^3}   e^{\phi/2}  4 C_K {\hat e}^{4567}   \wedge \big(   Re(g^2)_z e^{z23} + Re(g^2)_{\bar z} e^{\bar z 23}\big)\no\\
 &=& \int_{M_4\times S^3}   4  e^{\phi/2}  f_2^2 \rho C_K \big(   Re(g^2)_z dz+  Re(g^2)_{\bar z} d\bar z\big) \hat e^{234567}\no\\
 &=& Vol(S^2) \int 4  e^{\phi/2}  f_2^2 \rho C_K \big(   Re(g^2)_z dz+  Re(g^2)_{\bar z} d\bar z\big) \no\\
 &=& Vol(S^2) \int 4  C_K \big( \partial_{ z} b^{(2)} dz +\partial_{\bar z} b^{(2)} d{\bar z} \big)\no \\
 &=& Vol(S^2) \int \Big( -{2 i } {B^2-\bar B^2\over A+\bar A} -{2}\tilde K\Big)  \big( \partial_{ z} b^{(2)} dz +\partial_{\bar z} b^{(2)} d{\bar z} \big)\no\\
\eea

\subsubsection{Fundamental string charge}\label{pagef1ch}
The relevant terms in the F1 brane charge are given by
\bea\label{qf1page}
Q_{F1}^{Page } &=& - \int_{S^7} \Big( e^{-\phi} * H_3 -\chi e^\phi *\tilde F_3 +4 C_4\wedge dC_2)\no\\
&=&  -\int_{M_4 \times S^3} \Big( e^{-\phi} * H_3-\chi e^{\phi}  *\tilde F_3+4 C_4\wedge (\tilde F_3 +\chi H_3)\Big)
\eea
The three parts making up this charge are given by
\bea
 \int_{S^7}  e^{-\phi} * H_3 &=& Vol(S^2) \int e^{-\phi/2} f_3^4 f_2^2 \rho \Big( -i Re(g^1)_z dz + i Re(g^1)_{\bar z} d\bar z \Big) \no\\
 &=&  Vol(S^2)  \int e^{-\phi} { f_3^4 f_2^2 \over f_1^2} \Big(- i \partial_z b^{(1)} dz+ i  \partial_{\bar z} b^{(1)} d\bar z\Big) \no\\
 &=& Vol(S^2)  \int {\Big( (A+\bar A)K-(B+\bar B)^2\Big)^2 \over K(A+\bar A)}\Big( -i \partial_z b^{(1)} dz+ i  \partial_{\bar z} b^{(1)} d\bar z\Big)\no\\
\eea
 The second part is given by
\bea
- \int_{S^7}\chi e^\phi *\tilde F_3 &=& - Vol(S^2) \int e^{\phi/2} \chi f_3^4 f_2^2 \rho   \Big(- i Im(g^1)_z dz + i Im(g^1)_{\bar z} d\bar z \Big)  \no\\
&=&   Vol(S^2) \int e^{\phi} \chi {f_3^4 f_2^2 \over f_1^2}  \Big( i  ( \partial_z c^{(1)} -\chi \partial_z b^{(1)}) dz - i  ( \partial_{\bar z} c^{(1)} -\chi \partial_{\bar z} b^{(1)}) d\bar z\Big)\no\\
&=&   Vol(S^2)  \int   {4 K \over A+\bar A}  {(A+\bar A)K-(B+\bar B)^2\over (A+\bar A)K -(B-\bar B)^2}  \;  \no\\
&& \quad \times  \;  \chi \Big( i  ( \partial_z c^{(1)} -\chi \partial_z b^{(1)}) dz - i  ( \partial_{\bar z} c^{(1)} -\chi \partial_{\bar z} b^{(1)}) d\bar z\Big)
\eea
 The third term is given by
\bea\label{cfourb}
\int 4 C_4\wedge dC_2&=& \int 4C_4\wedge (\tilde F_3+ \chi H_3)\no \\
&=& \int 4C_4 \wedge \Big(e^{-\phi/2} Im(G) +\chi  e^{\phi/2}Re(G)\Big) \no\\
&=& Vol(S^2)\int 4 C_K \Big(   \partial_z c^{(2)} dz+ \partial_{\bar z} c^{(2)} d\bar z\Big)\no\\
&=& - Vol(S^2) \int \Big({2i } {B^2-\bar B^2\over A+\bar A} +{2  }\tilde K \Big) \Big(   \partial_z c^{(2)} dz+ \partial_{\bar z} c^{(2)} d\bar z\Big)
\eea

\section{Six Dimensional perspective}\label{sixdred}
\setcounter{equation}{0}

Type IIB supergravity on a $M_4$ manifold gives a six-dimensional $\mathcal{N}=(2,0)$ chiral theory
with 105 scalar and 26 tensor fields. For $M_4=K_3$,
the scalar fields live in a $SO(5,21)/SO(5) \times SO(21)$ coset, while the three-from anti-symmetric tensor fields transform as a vector under the action of the
$SO(5,21)$ global symmetry. The theory has a $SO(5)$ R-symmetry.
The coset can be parameterized using a vielbein \cite{Romans:1986er} $V^{a}_{A}$.  The vielbein is a $SO(5,21)$ matrix and satisfies
\be
\eta= V^{T} \eta V
\ee
where $\eta={\rm diag}({\bf 1}_{5} , -{\bf 1}_{21})$ is the $SO(5,21)$ invariant metric.
The field strength $H^{i}$ and $H^{r}$ are self dual and anti self dual respectively
  \bea
H^{i}_{\mu\nu\rho}&=& {\sqrt{-g}\over 3!} \epsilon_{\mu\nu\rho\lambda\sigma\tau} H^{i\; \lambda\sigma\tau} , \quad \quad i=1,2\cdots ,5\no\\
H^{r}_{\mu\nu\rho}&=&- {\sqrt{-g}\over 3!} \epsilon_{\mu\nu\rho\lambda\sigma\tau} H^{r\; \lambda\sigma\tau} , \quad \quad r=1,2\cdots ,21
\eea
where $g$ is the determinant of $g_{\mu\nu}$ and $\epsilon$ is the six-dimensional completely antisymmetric tensor which satisfies $\epsilon_{012345}=+1$.
The field strength $H$ does not satisfy a simple Bianchi identity.  They can be expressed using the vielbein as
\bea
H^{i}_{\mu\nu\rho} &=& V^{i}_{\;A} G^{A}_{\mu\nu\rho} \no\\
H^{r} _{\mu\nu\rho}&=& V^{r}_{\;A} G_{\mu\nu\rho} ^{A}
\eea
Where  $G^{A}$ satisfies a simple Bianchi identity, i.e.  $dG^{A}=0$, for all $A=1,2,\cdots, 26$.

In the solutions constructed in \cite{Chiodaroli:2009yw}, the internal moduli of the $M_4$ have constant profiles.
From a six-dimensional perspective, four scalars ($\phi$,
 $\chi$, the four-form potential in the compact directions $C_K$ and the metric factor $f_3$) and two tensor fields ($H_3$ and $F_3$) have non-trivial profiles
These non-trivial fields give a $SO(2,2)/SO(2)\times SO(2)$ coset with the self-dual and anti self-dual parts of $H_3$ and $F_3$ transforming as a ${\bf 4}$ with respect to
the global $SO(2,2)$ symmetry.\\
We can reduce the equations of motion (\ref{eqofast3}) to six-dimensions obtaining,
\bea d * \big( e^{- \phi} f^4_3 \tilde F_3 \big) &=& 4 d C_K \wedge H_3 \\
d * \big( e^{- \phi} f^4_3 H_3 -e^{\phi } f^4_3 \chi \tilde F_3 \big) &=& -4 d C_K \wedge F_3 \eea
The Bianchi identities for the three-form fields are unchanged,
\be dH_3 = 0, \qquad  d \tilde F_3 = - d\chi \wedge H_3 \ee
In order to obtain an explicit expression for the $SO(2,2)$ vielbein, we note that the following combinations of three-form tensor fields
obey to standard Bianchi identities,
\bea G_{(3)} &=& e^{-  \phi} f^4_3 *H_{(3)} - e^{\phi } f^4_3 \chi * \tilde F_{(3)} + 4 C_K F_{(3)} \\
E_{(3)}  &=& e^{\phi} f^4_3 * \tilde F_{(3)} -4 C_K H_{(3)} \\
 F_{(3)} &=& \tilde F_{(3)} + \chi H_{(3)}    \\
  H_{(3)} & &    \eea
We can then show that
\be \left( \begin{array}{c} *_6 G_{(3)} \\ *_6 E_{(3)} \\ *_6 H_{(3)} \\ *_6 F_{(3)}  \end{array} \right) = \eta_1 \; M \left( \begin{array}{c} G_{(3)} \\ E_{(3)} \\ H_{(3)} \\ F_{(3)}   \end{array}\right) \ee
The matrix $M$ is an element of $SO(2,2)$: $M^T \eta_1 M = M \eta_1 M^T = \eta_1$. The choice for $\eta_1$ is non-standard,
\be \eta_1 = \left( \begin{array}{cc} 0 & I_2 \\ I_2 & 0 \end{array} \right) \ee
It is possible to show that the matrix $M$ can be written as $\tilde V^T  \tilde V$ with
\be \tilde  V^a_{\;A} = \left( \begin{array}{cccc} e^{\phi \over 2} f^{-2}_3 &  e^{\phi \over 2 } f^{-2}_3 \chi & 4 e^{\phi \over 2} f^{-2}_3 \chi C_K & - 4 e^{\phi \over 2} f^{-2}_3 C_K \\
 0 & e^{ -{ \phi \over 2 }} f^{-2}_3 & 4 e^{ - {\phi \over 2}} f^{-2}_3 C_K & 0   \\
 0  & 0 & e^{ -  {\phi \over 2}} f^{2}_3  & 0 \\
0  & 0 &  - e^{\phi \over 2 }f^{2}_3 \chi  & e^{\phi \over 2}f^{2}_3  \end{array}
  \right) \ee
$\tilde V$ is an $SO(2,2)$ vielbein as expected,
\be \tilde V^T \eta_1  \tilde V = \eta_1, \quad \tilde  V \eta_1 \tilde V^T=\eta_1 \ee
We can use an orthogonal transformation to bring the $SO(2,2)$ invariant matrix in the standard form.
If we define
\be V= R^{-1} \tilde V R, \qquad  R= {1 \over \sqrt{2}} \left(  \begin{array}{cc} I_2 & -I_2 \\ I_2 & I_2  \end{array}  \right)
 \ee
With this definition, the vielbein obeys to
\be V^T \eta V = \eta, \quad V \eta V^T=\eta \ee
where the $SO(2,2)$ invariant metric is $ \eta = \text{diag}(+1,+1,-1,-1)$
Global $SO(2,2)$ transformations act on the right of $V$ as $V \rightarrow H V G$. Local $SO(2)\times SO(2)$ transformation act on the left as,
\be V \rightarrow H V, \qquad H= \left( \begin{array}{cc} S_1 & 0 \\0 & S_2 \end{array} \right) \ee
where $S_1$ and $S_2$ are $2 \times 2$ orthogonal matrices corresponding to the two $SO(2)$.
The vielbein $V$ can be used to construct self-dual and anti self-dual tensor combinations starting from three-form tensor fields obeying
to standard Bianchi identities,
\be H_{(3)}^I = V^I_A G_{(3)}^A, \qquad *_6 H_{(3)}^I = \eta^{I J} H_{(3)}^J, \qquad dG_{(3)}^A = 0     \ee
The charges associated to the tensor fields $G^A_{(3)}$ are exactly the Page charges:
\be Q^A = R^{-1} ( -Q^{Page}_{F1}, \; -Q^{Page}_{D1}, \; Q^{Page}_{NS5}, \; Q^{Page}_{D5})^T  \ee

\end{document}